\documentclass[aps,prd,twocolumn,showpacs,nofootinbib]{revtex4-1}

\usepackage{graphicx}
\usepackage{amsmath}
\usepackage{amssymb}
\usepackage{bm}
\usepackage{bbm}
\usepackage{color}
\usepackage{slashed}
\usepackage[colorlinks=true,linkcolor=blue,citecolor=blue, urlcolor=blue]{hyperref}

\begin{document}

\title{QCD NLO fragmentation functions for $c$ or $\bar{b}$ quark to $B_c$ or $B_c^*$ meson and their application}

\author{Xu-Chang Zheng}
\email{zhengxc@cqu.edu.cn}
\affiliation{Department of Physics, Chongqing University, Chongqing 401331, P.R. China.\\
Key Laboratory of Theoretical Physics, Institute of Theoretical Physics, Chinese Academy of Sciences, Beijing 100190, China.\\
School of Physical Sciences, University of Chinese Academy of Sciences, Beijing 100049, China.}
\author{Chao-Hsi Chang}
\email{zhangzx@itp.ac.cn}
\affiliation{Key Laboratory of Theoretical Physics, Institute of Theoretical Physics, Chinese Academy of Sciences, Beijing 100190, China.\\
School of Physical Sciences, University of Chinese Academy of Sciences, Beijing 100049, China.\\
CCAST (World Laboratory), Beijing 100190, China.}
\author{Tai-Fu Feng}
\email{fengtf@hbu.edu.cn}
\affiliation{Department of Physics, Hebei University, Baoding 071002, China.\\
Key Laboratory of Theoretical Physics, Institute of Theoretical Physics, Chinese Academy of Sciences, Beijing 100190, China.}
\author{Xing-Gang Wu}
\email{wuxg@cqu.edu.cn}
\affiliation{Department of Physics, Chongqing University, Chongqing 401331, P.R. China.}
\date{\today}

\begin{abstract}
The fragmentation functions for a $c$ or $\bar{b}$ quark to a $B_c$ or $B_c^*$ meson are derived up to QCD next-to-leading order. They are further computed numerically and presented  precisely in figures. In order to reach a higher accuracy, we also try to properly use them to estimate $B_c$ and $B_c^*$ production at a Z factory (an $e^+e^-$ collider running at the energy of the Z-boson pole).\\


\end{abstract}

\maketitle

\section{Introduction}

$B_c$ and $B_c^*$ are the ground states of the $c$,$\bar{b}$ binding system with spin $0$ and $1$ respectively. Carrying two different
heavy flavors, they are unique doubly heavy mesons in the Standard Model. Thus they attract a lot of attentions,  particularly after the $B_c$ meson was first observed\cite{CDF}. Their components, being of heavy flavor quarks, move
nonrelativistically inside the mesons, so the effective theory
—nonrelativistic quantum chromodynamics (NRQCD) \cite{nrqcd}
—is applicable, and the Mandelstam formulation of the
Bethe-Salpeter equation\cite{mand} under the instantaneous
approximation also works well.

The production of $B_c$ or $B_c^*$ in $e^+e^-$ collisions at the
Z-boson resonance [i.e., $e^+e^- \to Z/\gamma \to B_c(B_c^*)+X$]
under the framework of NRQCD or the Mandelstam formulation under the instantaneous approximation can be factorized as follows \cite{doublyhadron,ybook1,ybook2}:
\begin{eqnarray}\label{eqbcfact0}
&&d\sigma_{e^+e^-\to B_c+X}=\sum_n d\tilde{\sigma}_{e^+e^-\to (c\bar{b})[n]+X}\langle {\cal O} ^{B_c}(n)\rangle\nonumber \\
&&d\sigma_{e^+e^-\to B_c^*+X}=\sum_{n'} d\tilde{\sigma}_{e^+e^-\to (c\bar{b})[n']+X}\langle {\cal O} ^{B_c^*}(n')\rangle
\end{eqnarray}
where $d\tilde{\sigma}$ denotes the cross section for the perturbative production of the two-quark state $(c\bar{b})[n]$ [or $(c\bar{b})[n']$] with proper quantum numbers $n$ (or $n'$), which can be calculated using perturbative QCD (pQCD), and the nonperturbative matrix element $\langle {\cal O} ^{B_c}(n)\rangle$ [or $\langle {\cal O} ^{B^*_c}(n')\rangle$] representing the
transition probability from the perturbative two-quark state
$(c\bar{b})[n]$ [or $(c\bar{b})[n']$] into the hadronic state (a $B_c$ or $B^*_c$ meson) can be related to the wave function at origin of the $(c\bar{b})$ binding system squared in the potential model framework, and can also be calculated using lattice QCD.

Since the $B^*_c$ meson is similar to the $B_c$ meson [the
difference is that the spin of the diquark $(c\bar{b})$ inside $B_c$ is $S=0$ but the spin of the diquark $(c\bar{b})$ inside $B^*_c$ is $S=1$], throughout the paper we often use $B_c$ to represent both $B_c$ and $B_c^*$ for simplicity.

However, when the center-of-mass energy of a collision
is larger than the heavy-quark mass and the terms in
${\cal O}(m^2_Q/s)$ can be neglected, according to the factorization formulation of pQCD the production can also be calculated in terms of the fragmentation approach:
\begin{eqnarray}
\frac{d\sigma_{e^+e^-\to B_c+X}}{dz}=&&\sum_i \int_z^1 \frac{dy}{y}\frac{d\hat{\sigma}_{e^+e^-\to i+X}}{dy}(y,\mu_F) \nonumber \\
&&D_{i\to B_c}(z/y,\mu_F),\label{eqbcfact}
\end{eqnarray}
where $z\equiv 2 p \cdot q/q^2$ is the energy fraction (e.g. here $p$ is the momentum of $B_c$, and $q$ is the momentum of $e^+$ and $e^-$ collision), $d\hat{\sigma}_{e^+e^-\to i+X}$ is the cross section (coefficient function) for the inclusive production of a parton $i$($i=c, \bar{b}$, etc.) and can be calculated using pQCD, $\mu_F$ denotes the factorization scale for the production, and $D_{i\to B_c}$ is the fragmentation function (FF) from a parton $i$ to a $B_c$ meson, which is universal and can be extracted experimentally. The authors of Refs.\cite{fraglo1,fraglo2} realized that the production is calculable in terms of QCD factorization as shown in Eq.(\ref{eqbcfact0}) and the leading-order (LO) FFs can be extracted by comparing Eqs.(\ref{eqbcfact0}) and (\ref{eqbcfact}), i.e., the FFs are theoretical calculable, and they were first extracted in Refs.\cite{fraglo1,fraglo2}. The authors of Ref.\cite{comparlo} applied the obtained FFs to the production to the QCD leading logarithm approach and made comparisons between their results and those obtained using the complete LO QCD approach, which gives us a understanding of the two approaches.

In order to obtain a better theoretical estimation on $B_c$ production, etc., at a Z factory\cite{changch} (an $e^+e^-$ collider running at the energy of the Z-boson pole), we would like to adopt the factorization approach (\ref{eqbcfact}) but with the FFs from the $c$ or $\bar{b}$ quark to a $B_c$ meson which are of next-to-leading order (NLO), because the NLO QCD calculations are generally more accurate. The NLO FFs cannot be extracted from the complete NLO calculation of the relevant $B_c$ production as easily as those for the LO ones, although $B_c$ production at a Z factory has been studied using the ``complete computation approach"\cite{bcnlo}. Therefore we must start with the definition given in Ref.\cite{Collins} to derive them up to the NLO of QCD. In addition, the QCD NLO FFs have many applications, so we would like to derive them precisely here, although the derivation is complicated.

Note that in Refs.\cite{braaten,braaten1,YJia,YQMa} the QCD NLO FFs for a gluon to heavy quarkonium were derived, but here the FFs from a quark $i$ to a $B_c$ meson involve two heavy quarks of different flavors, so they are quite different from the ones for a gluon to heavy quarkonium.

According to NRQCD, the FFs $D_{i\to B_c}$ ($i=c,\bar{b}$), which
depict the hadronization and contain nonperturbative effects, can be factorized as follows:
\begin{eqnarray}
D_{i \to B_c}(z,\mu_F)=\sum_n d_{i\to c\bar{b}[n]}(z,\mu_F)\langle {\cal O}_n^{B_c} \rangle,\label{ffnrqcd}
\end{eqnarray}
where the first factor $d_{i\to c\bar{b}[n]}$ denotes a parton $i$ generating a $c\bar{b}$ quark pair with matched quantum number $n$, and being perturbative it can be calculated using pQCD; the factor $\langle {\cal O}_n^{B_c} \rangle$ denotes the ``long-distance matrix elements", and being nonperturbative they may be related to the wave functions at the origin in the potential model framework or computed using lattice QCD. The nonperturbative factors are reduced to a few long-distance matrix elements $\langle {\cal O}_n^{B_c} \rangle$ under the required accuracy\footnote{The relevant discussions about the accuracy of applying NRQCD to the FFs of heavy quarkonia can be found in Refs.\cite{fragnrqcd1,fragnrqcd2}, and the conclusions also apply to the FFs of the $B_c$ meson.}. With the normalization $\int_0^1 dz D_{i \to B_c(B^*_c)}(z)=1$, the LO FFs $D_{i \to B_c}$(where $i=c,\bar{b}$) were first obtained in Ref.\cite{fraglo1}. The LO FFs were extracted from the LO calculations of the processes $Z \to B_c+b+\bar{c}$ and $Z \to B_c^*+b+\bar{c}$ with the approximation $m_{B_c}\ll m_{_Z}$. Subsequent calculations\cite{fraglo2,fraglo3} confirmed the results. The LO FFs for the production of the P-wave and D-wave excited states of the $B_c$ were calculated in Refs.\cite{fraglo4,fraglo5,fraglo6}. So
far there is no NLO calculation for the FFs $D_{i\to B_c}$ ($i=c,\bar{b}$). Thus, in the present paper, we devote ourselves to calculating the QCD NLO corrections to $D_{i \to B_c}$ (and $D_{i \to B^*_c}$).

Since the FFs $D_{i \to B_c}(z,\mu_F)$(where $\mu_F$ is the factorization energy) generally contain terms like ${\rm ln}(\mu_F/m_Q)$, in order to properly take into account the possible large-logarithm terms the FFs $D_{i \to B_c}(z,\mu_F)$[$\mu_{F}={\cal O}(\sqrt{s})$] will be obtained by solving the Dokshitzer-Gribov-Lipatov-Altarelli-Parisi (DGLAP) evolution equations \cite{dglap1,dglap2,dglap3} with the NLO QCD FFs $D_{i \to B_c}(z,\mu_{F0})$[$\mu_{F0}={\cal O}(m_Q)$] being the ``initial FFs",
\begin{eqnarray}
&&\frac{d}{d~{\rm ln}{\mu^2_F}}D_{i \to B_c}(z,\mu_F)\nonumber \\
&&=\frac{\alpha_s(\mu_F)}{2\pi}\sum_j \int_z^1 \frac{dy}{y}P_{ji}(y,\alpha_s(\mu_F)) D_{j \to B_c}(z/y,\mu_F),\nonumber\\
\label{eq.dglap}
\end{eqnarray}
where $P_{ji}(y,\alpha_s(\mu_F))$ are splitting functions for parton $i$ into parton $j$\footnote{In fact, here they are of LO.}:
\begin{eqnarray}
P_{qq}(y)&=&C_F\left[\frac{1+y^2}{(1-y)_+}+\frac{3}{2}\delta(1-y)\right],\nonumber \\
P_{gq}(y)&=&C_F\frac{1+(1-y)^2}{y},\nonumber \\
P_{qg}(y)&=&T_F\left[ y^2+(1-y)^2 \right],\nonumber \\
P_{gg}(y)&=&2C_A\left[\frac{y}{(1-y)_+}+\frac{1-y}{y}+y(1-y)\right]\nonumber \\
&&+\frac{1}{6}\delta(1-y)(11C_A-4n_fT_F)\,,
\label{eq.spfun1}
\end{eqnarray}
where $C_F=4/3, T_F=1/2, C_A=3$ for QCD and $P_{\bar{q}\bar{q}}$ is equal to $P_{qq}$. Note that in order to focus on the consequences of NLO QCD corrections for FFs, we restrict ourselves to evaluating the evolution of the FFs from $\mu_{F0}$ to $\mu_{F}$ only to leading-logarithm (LL) accuracy so that here the ``splitting functions" in Eq.(\ref{eq.spfun1}) are of leading order.

The paper is organized as follows. Following the
Introduction, in Sec.\ref{LOFrag} we present the definition of the
FFs which was given by Collins and Soper\cite{Collins}, and
with this definition we calculate the LO FFs for
$i \to B_c(B^*_c)+\cdots (i=\bar{b},c)$. In Sec.\ref{sec3} we describe the adopted method for calculating the virtual and real corrections to the FFs, and how to carry out the renormalization,
so as to obtain the ``initial FF" $D_{i \to B_c(B^*_c)}(z,\mu_{F0})$. Then, we present the numerical results for the FFs $D_{i \to B_c}$ and $D_{i \to B^*_c}$ up to QCD NLO. In Sec.\ref{sec4} we apply the obtained QCD NLO FFs to the production of $e^+e^-\to B_c(B^*_c)+\cdots$ at a Z factory and compare the results with those obtained from the complete QCD NLO calculations. Section \ref{conclusion} is devoted to discussions and a conclusion.

\section{The fragmentation functions}
\label{LOFrag}

\subsection{The definition of fragmentation functions}

\begin{widetext}

The FFs may be defined as the hadron matrix elements of
certain quark-field operators, and the light-cone coordinate
is conventionally adopted. In the light-cone coordinate a vector in d-dimensional space-time\footnote{In this work, we adopt dimensional regularization with $d=4-2\epsilon$ to regularize UV and IR divergences, and adopt the reading point prescription\cite{gamma5} to handle $\gamma_5$ in $d$ dimensions.} is represented as $V^{\mu}=(V^+,V^-,V_T)=((V^0+V^{d-1})/\sqrt{2},(V^0-V^{d-1})/\sqrt{2},V_T)$.
The gauge-invariant definition of the FFs for a quark $Q$ fragmenting into a hadron $H$ in $d=4-2\epsilon$-dimensional
space-time is\cite{Collins}
\begin{eqnarray}
D_{Q\to H}(z)=&&\frac{z^{d-3}}{2\pi}\sum_{X} \int dx^- e^{-iP^+ x^-/z} \nonumber \\
&&\times \frac{1}{N_c} {\rm Tr}_{\rm color}  \frac{1}{4} {\rm Tr}_{\rm Dirac} \left\lbrace \gamma^+ \langle 0 \vert \Psi(0)\bar{{\cal P}} {\rm exp}\left[ig_s \int_{0}^{\infty} dy^- A_a^+(0^+,y^-,0_T)t_a^T \right]\vert H(P^+,0_T)+X \rangle \right. \nonumber \\
&&\left. \times\langle H(P^+,0_T)+X\vert {\cal P} {\rm exp}\left[-ig_s \int_{x^-}^{\infty} dy^- A_a^+(0^+,y^-,0_T)t_a^T \right] \bar{\Psi}(x)\vert 0\rangle\right\rbrace,
\label{defrag1}
\end{eqnarray}
\end{widetext}
where $\Psi$ is the quark field and $A_a^{\mu}$ is the gluon field. ${\cal P}$ denotes path ordering, $t^a$ is the color matrix, $z$ is the longitudinal momentum fraction $z=P^+/K^+$, and $K$ is the momentum of the initial quark $Q$. The FFs
are defined in the reference frame where the hadron $H$ carries the momentum $P^{\mu}=(P^+,P^-=m_H^2/2P^+,0_T)$. It is convenient to introduce a light-like vector $n^\mu=(0,1,0_T)$ in the reference frame where the FFs are defined. Then, the plus component of a momentum p can be written as $p^+=p \cdot n$, and $z=P\cdot n/K \cdot n$.

The definition\cite{Collins} of FFs for an antiquark $\bar{Q}$ into a hadron $H$ is
\begin{widetext}
\begin{eqnarray}
D_{\overline{Q}\to H}(z)=&&\frac{z^{d-3}}{N_c \times 4\times 2\pi}\sum_{X} \int dx^- e^{-iP^+ x^-/z} \nonumber \\
&&\times \langle 0\vert \bar{\Psi}(0) \gamma^+  \bar{{\cal P}} {\rm exp}\left[-ig_s \int_{0}^{\infty} dy^- A_a^+(0^+,y^-,0_T)t^a \right]\vert H(P^+,0_T)+X \rangle \nonumber \\
&&\times \langle H(P^+,0_T)+X\vert {\cal P} {\rm exp}\left[ig_s \int_{x^-}^{\infty} dy^- A_a^+(0^+,y^-,0_T)t^a \right] \Psi(x)\vert 0\rangle.
\label{defrag2}
\end{eqnarray}
\end{widetext}
Given the Feynman rules and the definition of the FFs (\ref{defrag1})-(\ref{defrag2}), the relevant Feynman diagrams can be drawn. The part to the left of the cut line in the Feynman diagram corresponds to the right part of the definition, and the part to the right of the cut corresponds to the left part of the
definition (which is just the complex conjugate of the right part of the definition). Note that for the FFs of an antiquark into a hadron we have the following:
\begin{itemize}
\item The vertex for a gluon line attached to an eikonal line contributes a factor $ig_s n^{\mu}t^a_{ij}$, where $\mu$ and $a$ are the Lorentz index and color index of the gluon, respectively.
\item The eikonal propagator, which carries momentum $q$ flowing from the operator to the cut side, is $i\delta_{ij}/(q\cdot n+i\epsilon)$.
\item The cut of final-state eikonal line carrying momentum $q$ contributes $2\pi\delta(q\cdot n)$.
\end{itemize}
An overall factor of $N_{CS}=z^{1-2\epsilon}/8\pi N_c$ from the definition should also be taken into account. The Feynman rules of the FFs of a quark into a hadron are the same as those in the
antiquark cases except that the color matrix for the eikonal
line–gluon vertex should be $t^a_{ij}$ instead of $-t^a_{ji}$. Thus, given the Feynman diagrams the LO and NLO FFs $D_{i \to B_c}$ and $D_{i \to B^*_c}$ can be derived.

\begin{figure}[htbp]
\includegraphics[width=0.5\textwidth]{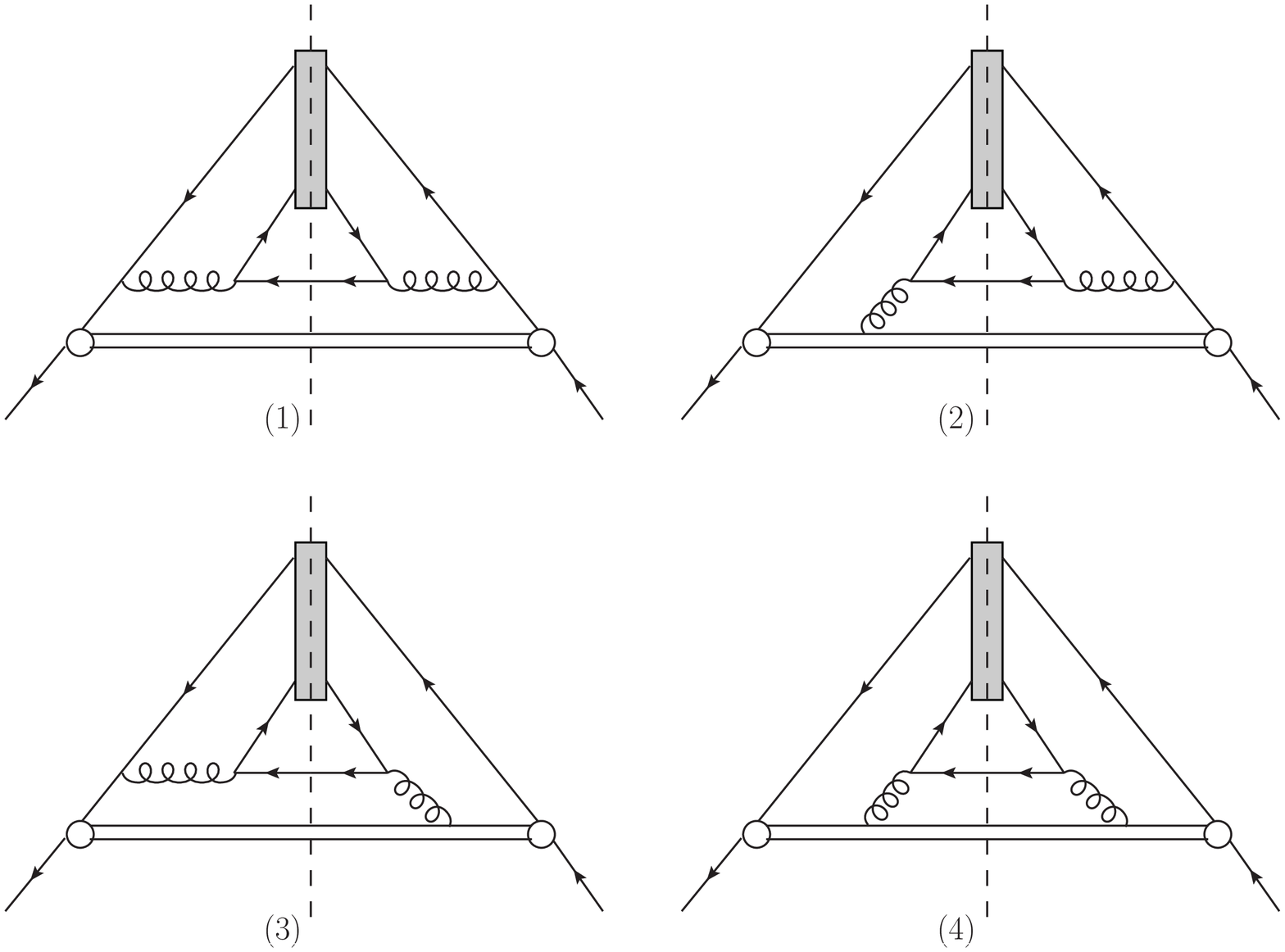}
\caption{The LO cut diagrams for the FFs $D_{\bar{b}\to c\bar{b}[n]}$. } \label{feylo}
\end{figure}

\subsection{LO fragmentation functions}

To understand the definition (\ref{defrag1})-(\ref{defrag2}) and to present the conventions used in this paper, here we derive the LO FFs, $D_{i \to B_c}$ and $D_{i \to B^*_c}$, where $i=c,\bar{b}$, although they have been obtained in the past using other approaches\cite{fraglo1,fraglo2,fraglo3}.

In this section and the next one we will show the derivations of the FFs $D_{\bar{b} \to B_c}$ and $D_{\bar{b} \to B^*_c}$ from the definition (\ref{defrag1})-(\ref{defrag2}). The FFs $D_{c \to B_c}$ and $D_{c \to B^*_c}$ can be derived in the same way and the results are the same as those for $D_{\bar{b} \to B_c}$ and $D_{\bar{b} \to B^*_c}$ with the replacement $m_b\leftrightarrow m_c$, so we will not repeat the derivation for them.

According to the factorization (\ref{ffnrqcd}), as the first step we derive the ``FFs" $D_{\bar{b}\to c\bar{b}[n]}$ with the diquark $c\bar{b}$ states with quantum numbers $^1S^{[1]}_0$ and $^3S^{[1]}_1$, where the superscript $[1]$ denotes the color singlet. Then the second step is to derive the FFs for a heavy quark ($\bar{b}$ or $c$) into a $B_c$ or $B^*_c$ meson, where the ``free diquark" $c\bar{b}$ state is replaced by the NRQCD matrix element (the wave function at the origin), which depicts QCD nonperturbative effects in the formation of a $B_c$ or $B^*_c$ meson from the relevant diquark state $c\bar{b}[n]$. (In this paper we assume that the QCD NLO matrix element is the same as the QCD LO one.\footnote{The matrix element appears as an overall factor, so its correction(s) (if any) can be considered easily.})

Based on the definition (\ref{defrag1})-(\ref{defrag2}), there are four cut diagrams (Fig.\ref{feylo}) for the LO FF $D_{\bar{b} \to c\bar{b}[n]}$. The squared Feynman amplitudes, corresponding to the four diagrams with a ``cut", can be written as follows:
\begin{eqnarray}
{\cal A}_1=&& {\rm tr}\left[\slashed{n} \frac{i}{-\slashed{p_1}-\slashed{p_2}-m_b+i\epsilon} (ig_s\gamma^{\mu}t^a)\Pi \right. \nonumber \\
&& \left. \cdot \Lambda_1(ig_s\gamma_{\mu}t^a)(\slashed{p}_{2}-m_c)(-ig_s\gamma^{\nu}t^b)\bar{\Pi} \right. \nonumber \\
&& \left. \cdot \Lambda_1 (-ig_s\gamma_{\nu}t^b) \frac{-i}{-\slashed{p_1}-\slashed{p_2}-m_b-i\epsilon}\right] \nonumber \\
&& \cdot \frac{-i}{(p_{11}+p_2)^2+i\epsilon} \frac{i}{(p_{11}+p_2)^2-i\epsilon}\vert_{q=0},\\
{\cal A}_2=&& {\rm tr}\left[\slashed{n}(ig_s n^{\mu}t^a) \frac{i}{(p_{11}+p_2)\cdot n+i\epsilon}\Pi\right. \nonumber \\
&& \left. \cdot \Lambda_1 (ig_s\gamma_{\mu}t^a)(\slashed{p}_{2}-m_c)(-ig_s\gamma^{\nu}t^b)\bar{\Pi} \right. \nonumber \\
&& \left. \cdot \Lambda_1 (-ig_s\gamma_{\nu}t^b) \frac{-i}{-\slashed{p_1}-\slashed{p_2}-m_b-i\epsilon}\right] \nonumber \\
&& \cdot \frac{-i}{(p_{11}+p_2)^2+i\epsilon} \frac{i}{(p_{11}+p_2)^2-i\epsilon}\vert_{q=0},\\
{\cal A}_3=&& {\rm tr}\left[\slashed{n} \frac{i}{-\slashed{p_1}-\slashed{p_2}-m_b+i\epsilon}(ig_s\gamma^{\mu}t^a)\Pi \right. \nonumber \\
&& \left. \cdot \Lambda_1 (ig_s\gamma_{\mu}t^a)(\slashed{p}_{2}-m_c)(-ig_s\gamma^{\nu}t^b)\bar{\Pi}\right. \nonumber \\
&& \left. \cdot \Lambda_1 \frac{-i}{(p_{11}+p_2)\cdot n-i\epsilon} (-ig_s n_{\nu}t^b)\right] \nonumber \\
&& \cdot \frac{-i}{(p_{11}+p_2)^2+i\epsilon} \frac{i}{(p_{11}+p_2)^2-i\epsilon}\vert_{q=0},\\
{\cal A}_4=&& {\rm tr}\left[\slashed{n}(ig_s n^{\mu}t^a) \frac{i}{(p_{11}+p_2)\cdot n+i\epsilon}\Pi\right. \nonumber \\
&& \left. \cdot \Lambda_1 (ig_s\gamma_{\mu}t^a)(\slashed{p}_{2}-m_c)(-ig_s\gamma^{\nu}t^b)\bar{\Pi}\right. \nonumber \\
&& \left. \cdot \Lambda_1   \frac{-i}{(p_{11}+p_2)\cdot n-i\epsilon} (-ig_s n_{\nu}t^b)\right] \nonumber \\
&& \cdot \frac{-i}{(p_{11}+p_2)^2+i\epsilon} \frac{i}{(p_{11}+p_2)^2-i\epsilon} \vert_{q=0},
\end{eqnarray}
where $p_{11}$ and $p_{12}$ are the momenta of the $c$ quark and $\bar{b}$ quark inside the $c\bar{b}$ pair and
\begin{eqnarray}
p_{11}=\frac{m_c}{M}p_1+q,~~~~ p_{12}=\frac{m_b}{M}p_1-q,
\end{eqnarray}
where $M\approx m_b+m_c$ is the mass of the $c\bar{b}$ pair.
$\Pi$ is the spin projector: for the spin singlet it is
\begin{eqnarray}
\Pi= \frac{-\sqrt{M}}{{4{m_b}{m_c}}}(\slashed{p}_{12}- m_b) \gamma_5 (\slashed{p}_{11} + m_c)
\end{eqnarray}
and for the spin triplet it is
\begin{eqnarray}
\Pi= \frac{-\sqrt{M}}{{4{m_b}{m_c}}}(\slashed{p}_{12}- m_b) \slashed{\epsilon}(p_1) (\slashed{p}_{11} + m_c)
\end{eqnarray}
$\bar{\Pi}$ is defined as $\bar{\Pi}=\gamma^0 \Pi^{\dagger} \gamma^0$. The color-singlet projector is
\begin{eqnarray}
\Lambda_1= \frac{1}{\sqrt{3}} \textbf{1},
\end{eqnarray}
where $\textbf{1}$ is the unit matrix of the color $SU_c(3)$ group. Note that throughout the paper we work in the Feynman gauge.

Having taken traces, the squared amplitudes corresponding to the LO FFs can be written as follows:
\begin{eqnarray}
{\cal A}_{\rm LO}&=&\sum_{i=1}^{4}{\cal A}_i \nonumber \\
&=&\frac{C_F^2 g_s^4  K\cdot n}{r_c^2 z^2 (1-r_b z)^2 M}\sum_{i=2}^4 \frac{a_i M^{2(i-2)}}{(s_1-m_b^2)^i},\label{aborn}
\end{eqnarray}
where $r_c=m_c/M$ and $r_b=m_b/M$. $s_1=(p_1+p_2)^2$ is the invariant mass of the lowest (LO) final states $(c\bar{b}+\bar{c})$. The coefficients $a_i$ can be found in the Appendix \ref{Ap.coeai}.

The differential phase space for the LO FFs can be written as
\begin{eqnarray}
d\phi_{\rm LO}=\frac{\theta(p_2^+)dp_2^+}{4\pi p_2^+}\frac{ d^{d-2}\textbf{p}_{2\perp}}{(2\pi)^{d-2}}2\pi \delta(K^+-p_1^+-p_2^+),
\end{eqnarray}
where the $\delta$ function comes from the cut through the
eikonal line. The integration over $p_2^+$ can be carried out
precisely due to the $\delta$ function. The integrand does not
depend on the angles of $\textbf{p}_{2\perp}$, so the integration over the angles of $\textbf{p}_{2\perp}$ is trivial, and can be carried out too. Thus, now the differential phase space is reduced to
\begin{eqnarray}
d\phi_{\rm LO}=&& \frac{z^{-1+\epsilon}(1-z)^{-\epsilon}}{2(4\pi)^{1-\epsilon}\Gamma(1-\epsilon)K\cdot n} \nonumber \\
&&\times \left(s_1-\frac{M^2}{z}-\frac{m_c^2}{1-z}\right)^{-\epsilon}ds_1.\label{phslo}
\end{eqnarray}
The range of $s_1$ is from $(M^2/z+m_c^2/(1-z))$ to $\infty$. The LO FFs can be represented as
\begin{eqnarray}
D^{\rm LO}_{\bar{b}\to c\bar{b}[n]}(z)=N_{CS}\int d\phi_{\rm LO} {\cal A}_{\rm LO}.
\label{eqlo}
\end{eqnarray}
The integration over $s_1$ can be carried out with Eq.(\ref{aborn}). Integrating over $s_1$, we obtain
\begin{eqnarray}
&&D^{\rm LO}_{\bar{b}\to c\bar{b}[n]+\cdots}(z) \nonumber \\
&&=\frac{C_F^2\alpha_s^2(1-z)(4\pi)^{\epsilon}\Gamma(1+\epsilon)}{4N_c r_c^2z(1-r_b~z)^{4+2\epsilon}M^{3+2\epsilon}}\left[a_2+a_3\frac{(1+\epsilon)z(1-z)}{2(1-r_b~z)^2} \right. \nonumber \\
&&~~\left. +a_4\frac{(2+\epsilon)(1+\epsilon)z^2(1-z)^2}{6(1-r_b~z)^4}\right].
\end{eqnarray}
Setting $d=4$, we obtain
\begin{eqnarray}
&&D^{\rm LO}_{\bar{b}\to c\bar{b}[^1S_0^{[1]}]+\cdots}(z) \nonumber \\
&&=\frac{8\alpha_s^2 z(1-z)^2}{81r_c^2(1-r_b z)^6 M^3}[6-18(1-2r_c)z  \nonumber \\
&&~~ +(21-74r_c+68r_c^2)z^2-2r_b(6-19r_c+18r_c^2)z^3\nonumber \\
&&~~+3r_b^2(1-2r_c+2r_c^2)z^4]\frac{\langle {\cal O} ^{c\bar{b}[^1S_0^{[1]}]}(^1S_0^{[1]})\rangle}{2N_c},\label{lobc}
\end{eqnarray}
and
\begin{eqnarray}
&&D^{\rm LO}_{\bar{b}\to c\bar{b}[^3S_1^{[1]}]+\cdots}(z)\nonumber \\
&&= \frac{8\alpha_s^2 z(1-z)^2}{27r_c^2(1-r_b z)^6 M^3}[2-2(3-2r_c)z \nonumber \\
&&~~ +3(3-2r_c+4r_c^2)z^2-2r_b(4-r_c+2r_c^2)z^3\nonumber \\
&&~~+r_b^2(3-2r_c+2r_c^2)z^4]\frac{\langle {\cal O} ^{c\bar{b}[^3S_1^{[1]}]}(^3S_1^{[1]})\rangle}{6N_c}\label{lobc*},
\end{eqnarray}
where the LO FFs for the $(c\bar{b})$ states have been
written in the factorization form, and at order $\alpha_s^0$,
\begin{eqnarray}
&&\langle {\cal O} ^{c\bar{b}[^1S_0^{[1]}]}(^1S_0^{[1]})\rangle=2N_c, \nonumber \\
&&\langle {\cal O} ^{c\bar{b}[^3S_1^{[1]}]}(^3S_1^{[1]})\rangle=2(d-1)N_c
\end{eqnarray}
with the normalization for the NRQCD matrix elements as that in Ref.\cite{nrqcd}. $\langle {\cal O} ^{c\bar{b}[^1S_0^{[1]}]}(^1S_0^{[1]})\rangle$ and $\langle {\cal O} ^{c\bar{b}[^3S_1^{[1]}]}(^3S_1^{[1]})\rangle$ denote the NRQCD matrix elements for the states $(c\bar{b})$.

Thus, the LO FFs for the $B_c$ and $B_c^*$ mesons are obtained by replacing $\langle {\cal O} ^{c\bar{b}[^1S_0^{[1]}]}(^1S_0^{[1]})\rangle$ and $\langle {\cal O} ^{c\bar{b}[^3S_1^{[1]}]}(^3S_1^{[1]})\rangle$ with $\langle {\cal O} ^{B_c}(^1S_0^{[1]})\rangle$ and $\langle {\cal O} ^{B^*_c}(^3S_1^{[1]})\rangle$, respectively. The NRQCD matrix elements $\langle {\cal O} ^{B_c}(^1S_0^{[1]})\rangle$ and $\langle {\cal O} ^{B^*_c}(^3S_1^{[1]})\rangle$ can be estimated as follows:
\begin{eqnarray}
&&\langle {\cal O} ^{B_c}(^1S_0^{[1]})\rangle \approx N_c\vert R_S(0)\vert^2/(2\pi), \nonumber \\
&&\langle {\cal O} ^{B^*_c}(^3S_1^{[1]})\rangle \approx (d-1)N_c\vert R_S(0)\vert^2/(2\pi), \label{ldme}
\end{eqnarray}
where $R_S(0)$ is the radial wave function at the origin for the $B_c(B_c^*)$ meson. Replacing the NRQCD matrix elements in Eqs.(\ref{lobc}) and (\ref{lobc*}) with the NRQCD matrix elements in Eq.(\ref{ldme}), we obtain
\begin{eqnarray}
&&D^{\rm LO}_{\bar{b}\to B_c}(z) \nonumber \\
&&=\frac{2\alpha_s^2 z(1-z)^2\vert R_S(0) \vert^2}{81\pi r_c^2(1-r_b z)^6 M^3}[6-18(1-2r_c)z  \nonumber \\
&&~~ +(21-74r_c+68r_c^2)z^2-2r_b(6-19r_c+18r_c^2)z^3\nonumber \\
&&~~+3r_b^2(1-2r_c+2r_c^2)z^4],
\end{eqnarray}
and
\begin{eqnarray}
&&D^{\rm LO}_{\bar{b}\to B^*_c}(z)\nonumber \\
&&= \frac{2\alpha_s^2 z(1-z)^2\vert R_S(0) \vert^2}{27\pi r_c^2(1-r_b z)^6 M^3}[2-2(3-2r_c)z \nonumber \\
&&~~ +3(3-2r_c+4r_c^2)z^2-2r_b(4-r_c+2r_c^2)z^3\nonumber \\
&&~~+r_b^2(3-2r_c+2r_c^2)z^4].
\end{eqnarray}
The LO FFs $D^{\rm LO}_{\bar{b}\to B_c}(z)$ and $D^{\rm LO}_{\bar{b}\to B^*_c}(z)$ obtained here are exactly the same as those obtained in Refs.\cite{fraglo1,fraglo2}, although the authors of Refs.\cite{fraglo1,fraglo2} derived them in a different way.

\section{QCD NLO corrections to the FFs for a $\bar{b}$ quark to a  $B_c$ or $B_c^*$ meson}
\label{sec3}
In this section we will derive the NLO FFs as defined by Eqs.(\ref{defrag1}) and (\ref{defrag2}), and divide the derivation of the NLO corrections into virtual corrections, real corrections and renormalization for convenience. Finally we will compute them numerically and present them in figures.

\subsection{The virtual NLO corrections}

\begin{figure}[htbp]
\includegraphics[width=0.5\textwidth]{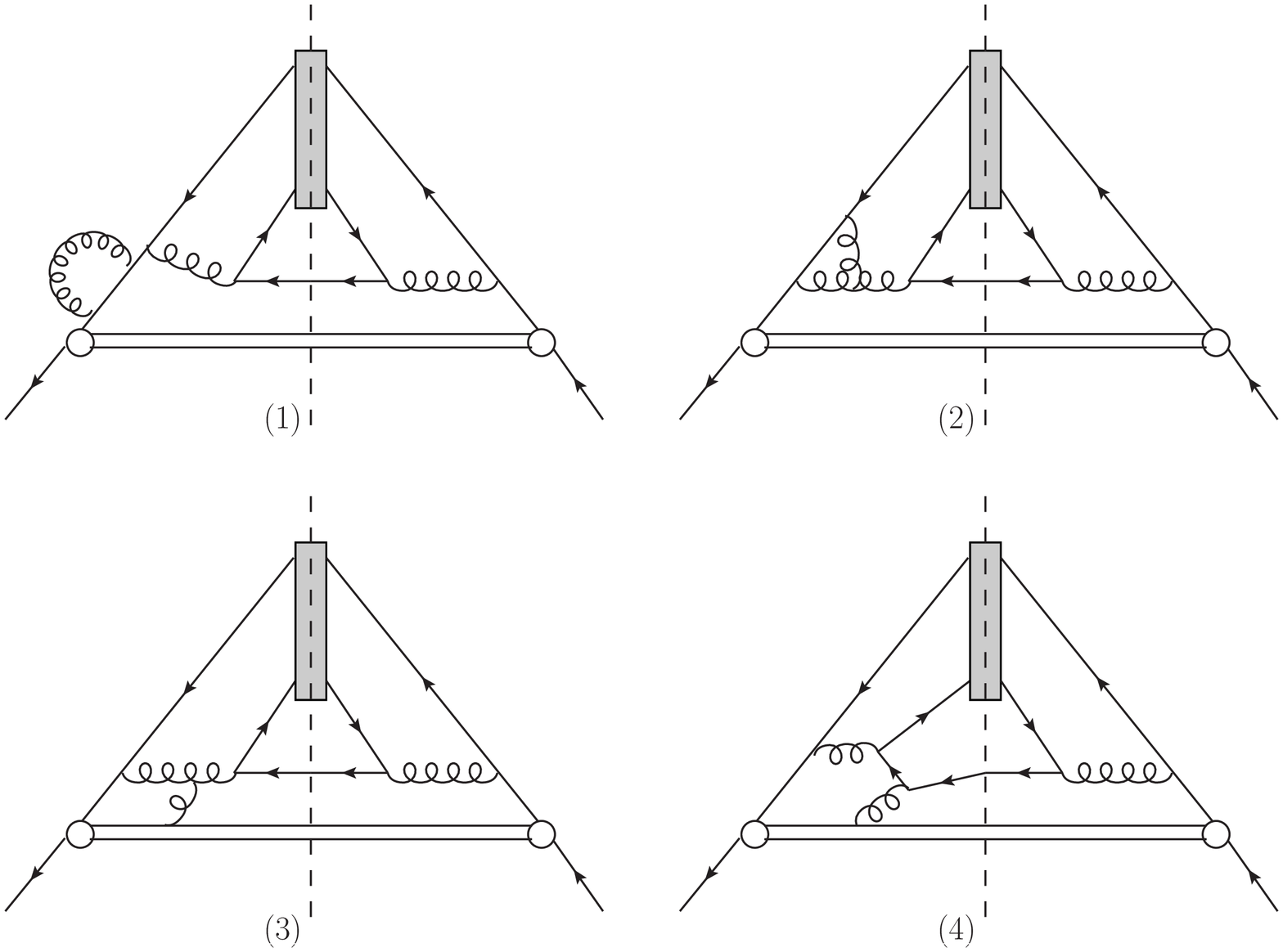}
\caption{Four sample cut diagrams for the virtual corrections to the FFs $D_{\bar{b}\to c\bar{b}[n]}$. } \label{feyvir}
\end{figure}

The virtual NLO corrections to the FFs $D_{\bar{b}\to c\bar{b}[n]}$ come from the ``cut diagrams" with one loop on either side of the cut. Four typical cut diagrams for the virtual corrections are shown in Fig.\ref{feyvir}.

There are Coulomb divergences in the conventional matching procedure. These Coulomb divergences may be regularized by a small relative velocity $v$ between the $\bar{b}$ quark and $c$ quark inside the produced $c\bar{b}$ pair. The Coulomb divergences also appear in the virtual corrections to the NRQCD matrix elements $\langle {\cal O} ^{c\bar{b}[^1S_0^{[1]}]}(^1S_0^{[1]})\rangle$ and $\langle {\cal O} ^{c\bar{b}[^3S_1^{[1]}]}(^3S_1^{[1]})\rangle$, while the NRQCD short-distance coefficients are free from Coulomb divergences at all. However, in dimensional regularization, we can avoid the divergence and extract the NRQCD short-distance coefficients by using the so-called region method\cite{region}. In the method, one may calculate the contributions from the hard region directly by expanding the relative momentum $q$ of the $c\bar{b}$ pair before performing the loop integration, and under the lowest nonrelativistic approximation one just needs to take $q=0$ before the loop integration. Thus the Coulomb divergences, which come from the potential region, do not appear in the calculations of the FFs for the free $c\bar{b}$ states and the NRQCD matrix elements. With this method, the NRQCD matrix elements $\langle {\cal O} ^{c\bar{b}[^1S_0^{[1]}]}(^1S_0^{[1]})\rangle$ and $\langle {\cal O} ^{c\bar{b}[^3S_1^{[1]}]}(^3S_1^{[1]})\rangle$ at NLO are the same as the LO ones.

The squared amplitudes of the virtual corrections can be read off from the virtual-correction cut diagrams with the Feynman rules in Section \ref{LOFrag}. The Dirac traces are carried out using the Mathmatica packages FeynCalc\cite{feyncalc1,feyncalc2} and FeynCalcFormlink\cite{formlink}. Then \$Apart\cite{apart} and FIRE\cite{fire} are adopted to do the partial fraction and integration-by-parts (IBP) reduction. After the IBP reduction, all one-loop integrals in amplitudes are reduced to master integrals. The master integrals include the common scalar one-loop integrals($A_0$, $B_0$ and $C_0$ functions) and the scalar one-loop integrals with one eikonal propagator. The $A_0$, $B_0$ and $C_0$ functions are calculated numerically using LoopTools\cite{looptools}. The scalar one-loop integrals with one eikonal propagator can be calculated using the method introduced in the Appendix of Ref.\cite{braaten}.

The differential phase space for the virtual corrections is the same as that for the LO FFs. The virtual corrections to the FFs can be expressed as
\begin{eqnarray}
D^{\rm virtual}_{\bar{b}\to c\bar{b}[n]}(z)=N_{CS}\int d\phi_{\rm LO} {\cal A}_{\rm virtual},
\label{eqvir}
\end{eqnarray}
where ${\cal A}_{\rm virtual}$ denotes the squared amplitudes for the virtual corrections.

\subsection{The real NLO corrections}

\begin{figure}[htbp]
\includegraphics[width=0.5\textwidth]{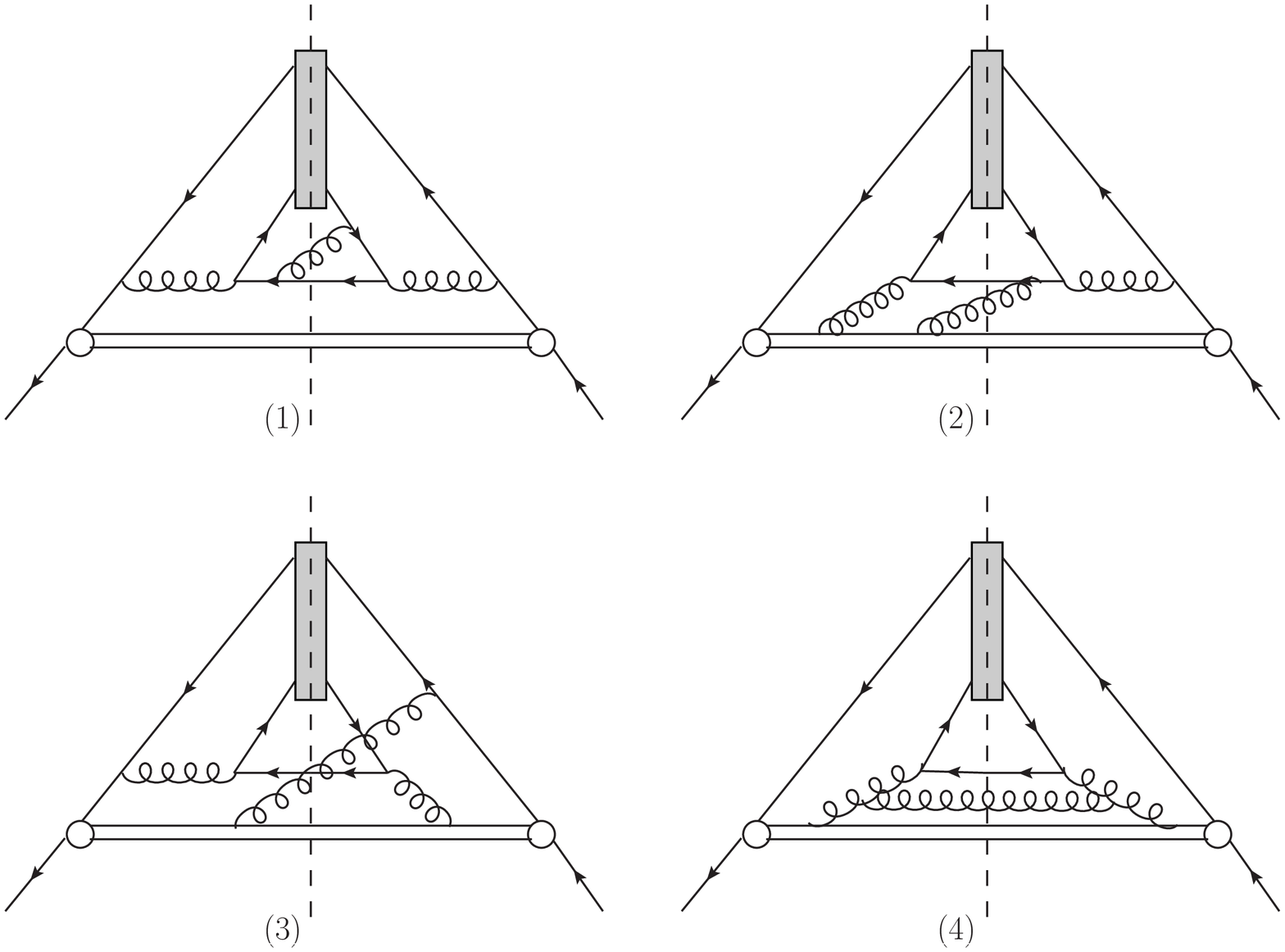}
\caption{Four sample cut diagrams for the real corrections to the FFs $D_{\bar{b}\to c\bar{b}[n]}$. } \label{feyreal}
\end{figure}

The real corrections to the FFs $D_{\bar{b}\to c\bar{b}[n]}$ come from the fragmentation processes in which an additional gluon is emitted in comparison with the corresponding LO ones. We denote the momenta of the initial and final particles as $\bar{b}(K)\to c\bar{b}[n](p_1)+\bar{c}(p_2)+g(p_3)$. The cut diagrams can be obtained from the LO cut diagrams in Fig.\ref{feylo} by adding a gluon line crossing the cut and connecting two of the lines on each side of the cut. Four typical cut diagrams for the real corrections are shown in Fig.\ref{feyreal}.

The differential phase space for the real corrections to the FFs can be written as
\begin{eqnarray}
d\phi_{\rm real}=&&2\pi \delta(K^+-p_1^+-p_2^+-p_3^+) \nonumber \\
&&\times \prod_{i=2,3}\frac{\theta(p_i^+)dp_i^+}{4\pi p_i^+}\frac{ d^{d-2}\textbf{p}_{i\perp}}{(2\pi)^{d-2}}.
\end{eqnarray}
The real corrections to the FFs can be written as
\begin{eqnarray}
D^{\rm real}_{\bar{b}\to c\bar{b}[n]}(z)=N_{CS}\int d\phi_{\rm real} {\cal A}_{\rm real},
\label{eqreal}
\end{eqnarray}
where ${\cal A}_{\rm real}$ denotes the squared amplitudes for the real corrections.

There are UV and IR divergences in the real corrections. These divergences come from the phase-space integration over the momentum of the final gluon $p_3$, and yield UV and IR poles in $\epsilon$ in dimensional regularization. However, it is impractical to do the phase-space integration for ${\cal A}_{\rm real}$ analytically. We follow the strategy used in Ref.\cite{braaten} to calculate the real corrections to the FF $D_{g\to \eta_Q}$, in order to extract the UV and IR poles. Namely,  we construct the subtraction terms ${\cal A}_S$ which have the same singularities as ${\cal A}_{\rm real}$ in the phase space, but the subtraction terms are simpler than ${\cal A}_{\rm real}$, and can be analytically integrated out over the phase space. Then the real corrections can be expressed as
\begin{eqnarray}
&&D^{\rm real}_{\bar{b}\to c\bar{b}[n]}(z)=N_{CS}\int d\phi_{\rm real} ({\cal A}_{\rm real}-{\cal A}_S)\nonumber \\
&&\;\;\;\;\;\;\;\;+N_{CS}\int d\phi_{\rm real} {\cal A}_S. \label{dsub}
\end{eqnarray}
Therefore, the first term on the right-hand side of Eq.(\ref{dsub}) is finite and can be calculated directly in four-dimensional space-time.

The UV divergences in the real corrections arise from the integrations over the phase-space region $p_{3\perp}\to \infty$. The IR divergences arise from the regions $p^+_{3}\to 0$ and $p_{3}\to 0$. The squared amplitudes for the real corrections can be expressed as
\begin{widetext}
\begin{eqnarray}\label{dsub1}
{\cal A}_{\rm real}=&&\frac{b_1(s_1,z)}{(1-y)(s-m_b^2)}+\frac{b_2(s_1,z)}{(1-y)(s_2-m_b^2)}+\frac{b_3(s_1,z)}{(1-y)s_3}+\frac{c_1(s_1,z,y)}{s-m_b^2}+\frac{c_2(s_1,z,y)p_1\cdot p_3}{(s-m_b^2)^2}\nonumber \\
&&+\frac{c_3(s_1,z,y)}{s_2-m_b^2}+\frac{c_4(s_1,z,y)p_2\cdot p_3}{(s_2-m_b^2)^2} +\frac{c_5(s_1,z,y)}{s_3}+\frac{c_6(s_1,z,y)p_1\cdot p_3}{s_3^2}+\frac{d_1(s_1,z)(1-u)(s_1-m_b^2)}{u~t_1(s-m_b^2)}\nonumber \\
&&+\frac{d_2(s_1,z)r_c(1-u)(s_1-m_b^2)}{u~t_1~s_3}+\frac{d_3(s_1,z)r_c(1-u)(s_1-m_b^2)^2}{u~t_1~s_3(s-m_b^2)}+\frac{d_4(s_1,z)r_c(s_1-m_b^2)^2}{u~t_2 (s-m_b^2)s_3}+\frac{d_5(s_1,z)(s_1-m_b^2)}{u~t_2 (s-m_b^2)}\nonumber \\
&&+\frac{d_6(s_1,z)r_c(s_1-m_b^2)}{u~t_2~s_3}+\frac{g(s_1,z)r_c(s_1-m_b^2)^2}{u(s-m_b^2)s_3}+\frac{h(s_1,z)}{t_2^2}+{\cal A}^{\rm finite}_{\rm real},
\end{eqnarray}
\end{widetext}
where the Lorentz-invariant parameters are defined as follows:
\begin{eqnarray}
&&y=\frac{(p_1+p_2)\cdot n}{(p_1+p_2+p_3) \cdot n},\, u=\frac{p_3\cdot n}{(p_2+p_3) \cdot n}, \nonumber \\
&&s=(p_1+p_2+p_3)^2,~~~~ s_2=(p_{12}+p_3)^2, \nonumber \\
&&s_3=(p_{11}+p_2+p_3)^2,~~t_1=2 p_1 \cdot p_3 ,\nonumber \\
&&t_2=2 p_2 \cdot p_3.\label{defvar}
\end{eqnarray}
Since we only consider the production of color-singlet S-wave $c\bar{b}$ states, the $1/t_1^2$ and $1/t_1 t_2$ terms cancel in ${\cal A}_{\rm real}$\cite{bcnlo}. The coefficients $b_i, c_i, d_i, g$ and $h$ can be obtained from the squared real-correction amplitudes and the results are very lengthy, so we do not present them here. The integrals of the $b_i$ ($i=1,2,3$) terms are UV and IR divergent, and yield double poles $\frac{1}{\epsilon_{_{\rm UV}}} \frac{1}{\epsilon_{_{\rm IR}}}$ or $\frac{1}{\epsilon^2_{_{\rm IR}}}$. The integrals of the $c_i$ terms are UV divergent, and yield a UV pole $\frac{1}{\epsilon_{_{\rm UV}}}$. The integrals of the $d_i$ terms are IR divergent, and yield a double pole  $\frac{1}{\epsilon^2_{_{\rm IR}}}$. The integrals of the $g$ and $h$ terms are IR divergent, and yield an IR pole $\frac{1}{\epsilon_{_{\rm IR}}}$. The term ${\cal A}^{\rm finite}_{\rm real}$ represents the remaining terms in ${\cal A}_{\rm real}$ which do not contribute divergences.

Now the subtraction terms can be constructed as follows:
\begin{widetext}
\begin{eqnarray}\label{dsub2}
{\cal A}_{\rm S}=&&\frac{b_1(s_1,z)}{(1-y)(s-m_b^2)}+\frac{b_2(s_1,z)}{(1-y)(s_2-m_b^2)}+\frac{b_3(s_1,z)}{(1-y)s_3}+\frac{c_1(s_1,z,y)}{s} \nonumber \\
&&+\frac{c_2(s_1,z,y)}{s^2}\left[p_1\cdot p_3-\frac{z}{2y}\left(1-\frac{2}{y}\right)s_1-\frac{1-y}{2y}(s_1+(1-r_c^2)M^2)\right]+\frac{c_3(s_1,z,y)}{s_2}\nonumber \\
&&+\frac{c_4(s_1,z,y)}{s_2^2}\left[p_2\cdot p_3 +\frac{(y-z)M^2}{z}\left(\frac{r_b}{2}+\frac{1-y}{z}\right)-\frac{1-y}{2z}(s_1-(1+r_c^2)M^2)\right]+\frac{c_5(s_1,z,y)}{s_3}\nonumber \\
&&+\frac{c_6(s_1,z,y)}{s_3^2}\left[p_1\cdot p_3+\frac{r_c z(1-r_b z)-(1-y)(y-z)}{2(y-r_b z)^2}(s_1-m_b^2) \right]+\frac{d_1(\tilde{s},z)(1-u)(\tilde{s}-m_b^2)}{u~t_1(\tilde{s}-m_b^2+t_1/z)}\nonumber \\
&&+\frac{d_2(\tilde{s},z)(1-u)(\tilde{s}-m_b^2)}{u~t_1[\tilde{s}-m_b^2+(1-r_b z)t_1/(r_c z)]}+\frac{d_3(\tilde{s},z)(1-u)(\tilde{s}-m_b^2)^2}{u~t_1(\tilde{s}-m_b^2+t_1/z)[\tilde{s}-m_b^2+(1-r_b z)t_1/(r_c z)]}\nonumber \\
&&+\frac{d_4(\tilde{s},z)(\tilde{s}-m_b^2)^2}{u~t_2[\tilde{s}-m_b^2+t_2/(1-z)][\tilde{s}-m_b^2+(1-r_b z)t_2/(r_c(1-z))]}+\frac{d_5(\tilde{s},z)(\tilde{s}-m_b^2)}{u~t_2(\tilde{s}-m_b^2+t_2/(1-z))}\nonumber \\
&&+\frac{d_6(\tilde{s},z)(\tilde{s}-m_b^2)}{u~t_2[\tilde{s}-m_b^2+(1-r_b z)t_2/(r_c(1-z))]}+\frac{g(\tilde{s},z)(\tilde{s}-m_b^2)^2}{u[\tilde{s}-m_b^2+t_2/(1-z)][\tilde{s}-m_b^2+(1-r_b z)t_2/(r_c(1-z))]}\nonumber \\
&&+\frac{h(\tilde{s},z)}{t_2^2},
\end{eqnarray}
\end{widetext}
where $\tilde{s}$ is defined as
\begin{eqnarray}
\tilde{s}=(p_1+\tilde{p})^2,
\end{eqnarray}
where
\begin{eqnarray}
\tilde{p}^{\mu}=p_2^{\mu}+p_3^{\mu}-\frac{p_2\cdot p_3}{(p_2+p_3)\cdot n} n^{\mu}.
\end{eqnarray}
One can check that the integration of $({\cal A}_{\rm real}-{\cal A}_{\rm S})$ over the phase space is finite in four space-time dimensions.

In order to analytically extract the UV and IR poles in $\epsilon$ in the real corrections, it is better to choose proper phase-space parametrizations for the terms in ${\cal A}_{\rm S}$. Various phase-space parametrizations can be found in Appendix \ref{Ap.psr}.

To integrate the subtraction terms that contain $s$, we use the parametrization in Eq.(\ref{eqb10}) for the differential phase space. The expression of the differential phase space in Eq.(\ref{eqb10}) can be decomposed as
\begin{eqnarray}
N_{CS}d\phi_{\rm real}=N_{\rm LO}(p_1,p_2)d\phi_{\rm LO}(p_1,p_2)d\phi^{(3)}(p_1,p_2,p_3),\nonumber \\ \label{phsp1}
\end{eqnarray}
where the prefactor $N_{\rm LO}(p_1,p_2)$ is defined as
\begin{eqnarray}
N_{\rm LO}(p_1,p_2)=\frac{(z/y)^{1-2\epsilon}}{8\pi N_c},\label{phsp2}
\end{eqnarray}
and $d\phi_{\rm LO}(p_1,p_2)$ is defined as
\begin{eqnarray}
d\phi_{\rm LO}(p_1,p_2)=&&\frac{z^{-1+\epsilon}(y-z)^{-\epsilon}}{2(4\pi)^{1-\epsilon}\Gamma(1-\epsilon)K\cdot n}\nonumber \\
 &&\times \left(s_1-\frac{y}{z}M^2-\frac{y}{y-z}m_c^2\right)^{-\epsilon}ds_1,\label{phsp3}
\end{eqnarray}
$d\phi_{\rm LO}(p_1,p_2)$ represents the differential phase space for a $\bar{b}$ quark with longitudinal momentum $y K\cdot n$ to fragment into a $B_c$($B^*_c$) meson with longitudinal momentum $z K\cdot n$ at LO. $N_{\rm LO}(p_1,p_2)$ and $d\phi_{\rm LO}(p_1,p_2)$ reduce to $N_{\rm LO}$ and $d\phi_{\rm LO}$ respectively, if $y=1$. Then $d\phi^{(3)}(p_1,p_2,p_3)$ can be expressed as
\begin{eqnarray}
&& d\phi^{(3)}(p_1,p_2,p_3)\nonumber \\
&& =\frac{1}{4(2\pi)^{3-2\epsilon}}(s-s_1/y)^{-\epsilon}[y(1-y)]^{-\epsilon}ds\, dy\, d\Omega_{3\perp}.\label{phsp4}
\end{eqnarray}
The range of $y$ is from $z$ to 1, the range of $s_1$ is from $[M^2/(z/y)+y m_c^2/(y-z)]$ to $\infty$, and the range of $s$ is from $s_1/y$ to $\infty$.

Thus, with Eqs.(\ref{phsp1})-(\ref{phsp4}), we can obtain
\begin{eqnarray}
&&N_{CS}\int d\phi_{\rm real}\frac{c_1(s_1,z,y)}{s}\nonumber \\
&&=\frac{\Gamma(1+\epsilon)}{\epsilon\,(4\pi)^{2-\epsilon}}\int_z^1 dy (1-y)^{-\epsilon}\int Nd\phi_{\rm LO}(p_1,p_2) \nonumber \\
&& ~~~ \times c_1(s_1,z,y) s_1^{-\epsilon},
\end{eqnarray}
where $Nd\phi_{\rm LO}(p_1,p_2) \equiv N_{\rm LO}(p_1,p_2)d\phi_{\rm LO}(p_1,p_2)$. The remaining integral in this equation does not generate poles in $\epsilon$.

We can also obtain
\begin{eqnarray}
&&N_{CS}\int d\phi_{\rm real}\frac{b_1(s_1,z)}{(1-y)(s-m_b^2)}\nonumber \\
&&=\frac{\Gamma(1+\epsilon)}{\epsilon (4\pi)^{2-\epsilon}}\int_z^1 dy (1-y)^{-1-\epsilon} \int Nd\phi_{\rm LO}(p_1,p_2)\nonumber \\
&& ~~~ \times b_1(s_1,z) (s_1-ym_b^2)^{-\epsilon}.\label{eqint1}
\end{eqnarray}
The integration over $y$ will diverge if $\epsilon=0$ in the limit $y\to1$, and contribute an IR pole. In order to extract this IR pole, we use the plus prescription, where
\begin{eqnarray}
(1-y)^{-1+a\epsilon}=&&\frac{\delta(1-y)}{a\epsilon}+\sum_{n=0}^{\infty} \frac{(a\epsilon)^n}{n!}\left( \frac{{\rm ln}^n(1-y)}{1-y}\right)_+.\nonumber \\ \label{eqplus}
\end{eqnarray}

Inserting Eq.(\ref{eqplus}) into Eq.(\ref{eqint1}), we obtain
\begin{eqnarray}
&&N_{CS}\int d\phi_{\rm real}\frac{b_1(s_1,z)}{(1-y)(s-m_b^2)}\nonumber \\
&&=\frac{\Gamma(1+\epsilon)}{\epsilon (4\pi)^{2-\epsilon}}\left\{-\frac{1}{\epsilon}N_{CS}\int  d\phi_{\rm LO} b_1(s_1,z) (s_1-m_b^2)^{-\epsilon} \right.
\nonumber \\
&&~~~+\int_z^1 dy \left[\frac{1}{(1-y)_+}-\epsilon\left( \frac{{\rm ln}(1-y)}{1-y}\right)_+\right] \nonumber \\
&&~~~\times \left. \int Nd\phi_{\rm LO}(p_1,p_2) b_1(s_1,z) (s_1-ym_b^2)^{-\epsilon}\right\}\nonumber \\
&&~~~+{\cal O}(\epsilon),
\end{eqnarray}
where $d\phi_{\rm LO}$ is the LO differential phase space given by Eq.(\ref{phslo}).

The integration of the terms with the coefficient $c_2(s_1,z,y)$ over $\Omega_{3\perp}$ is not trivial. We first calculate the integration of the vector $p_3^{\mu}$ over $\Omega_{3\perp}$. According to Lorentz invariance, this integral can be expressed as
\begin{eqnarray}
\int p_3^{\mu} d\Omega_{3\perp}=A n^{\mu}+B(p_1+p_2)^{\mu}.\label{intp3}
\end{eqnarray}
We can determine the coefficients $A$ and $B$ by contracting both sides of Eq.(\ref{intp3}) with $n_{\mu}$ [and contracting with $(p_1+p_2)_{\mu}$]. Then we obtain
\begin{eqnarray}
A&=&\frac{\Omega_{\perp}}{2yK\cdot n}\left(s-\frac{2-y}{y}s_1\right),\label{intp3a}\\
B&=&\frac{1-y}{y}\Omega_{\perp},\label{intp3b}
\end{eqnarray}
where $\Omega_{\perp}$ is the total transverse solid angle and $\Omega_{\perp}= 2\pi^{1-\epsilon}/\Gamma(1-\epsilon)$.

Inserting Eqs.(\ref{intp3a})-(\ref{intp3b}) into Eq.(\ref{intp3}) and contracting both sides of Eq.(\ref{intp3}) with $p_{1\mu}$, we obtain
\begin{eqnarray}
&&\int d\Omega_{3\perp}\left[p_1\cdot p_3-\frac{z}{2y}\left(1-\frac{2}{y}\right)s_1\right.\nonumber\\
&&\left. -\frac{1-y}{2y}(s_1+(1-r_c^2)M^2)\right]=\frac{zs}{2y}\Omega_{\perp},
\end{eqnarray}
Carrying out the integration over $s$, we obtain
\begin{eqnarray}
&&N_{CS}\int d\phi_{\rm real}\frac{c_2(s_1,z,y)}{s^2} \nonumber \\
&&~~~ \times \left[p_1\cdot p_3-\frac{z}{2y}\left(1-\frac{2}{y}\right)s_1 -\frac{1-y}{2y}(s_1+(1-r_c^2)M^2)\right]\nonumber \\
&&=\frac{\Gamma(1+\epsilon)}{\epsilon (4\pi)^{2-\epsilon}}\int_z^1 dy (1-y)^{-\epsilon} \int Nd\phi_{\rm LO}(p_1,p_2) \nonumber \\
&& ~~~ \times c_2(s_1,z,y)(z/2y) s_1^{-\epsilon}.
\end{eqnarray}

The method used to extract the poles from the subtraction terms involving $s$ integration can also be used to extract the poles from the integrations over the $s_2$ and $s_3$ of the subtraction terms.

For the $s_2$ integration, we adopt the parametrization in Eq.(\ref{eqb13}). The expression in Eq.(\ref{eqb13}) can also be decomposed to the form of Eq.(\ref{phsp1}), but the expression for $d\phi^{(3)}(p_1,p_2,p_3)$ becomes
\begin{eqnarray}
&&d\phi^{(3)}(p_1,p_2,p_3)\nonumber \\
&&=\frac{1}{4(2\pi)^{3-2\epsilon}}(r_b z)^{-1+\epsilon}y^{1-2\epsilon}(1-y)^{-\epsilon}\nonumber \\
&&~~~ \times \left(s_2-\frac{1-y+r_bz}{r_bz}m_b^2\right)^{-\epsilon}dy\, ds_2\, d\Omega_{3\perp}.\label{phsp5}
\end{eqnarray}
The ranges of $y$ and $s_1$ are the same as above. The range of $s_2$ is from $(1-y+r_b z)m_b^2/r_bz$ to $\infty$. Then we can readily obtain
\begin{eqnarray}
&&N_{CS}\int d\phi_{\rm real}\frac{c_3(s_1,z,y)}{s_2}\nonumber \\
&&=\frac{\Gamma(1+\epsilon)(m_b^2)^{-\epsilon}}{\epsilon(4\pi)^{2-\epsilon}}\int_z^1 dy [(1-y)(1-y+r_b z)]^{-\epsilon}\left(\frac{y}{r_b z} \right)^{1-2\epsilon}\nonumber \\
&&~~~ \times \int Nd\phi_{\rm LO}(p_1,p_2) c_3(s_1,z,y),\\
&&N_{CS}\int d\phi_{\rm real}\frac{b_2(s_1,z)}{(1-y)(s_2-m_b^2)}\nonumber \\
&&=\frac{\Gamma(1+\epsilon)(m_b^2)^{-\epsilon}}{\epsilon(4\pi)^{2-\epsilon}}(r_b z)^{-1+2\epsilon}\left\{-\frac{1}{2\epsilon}N_{CS}\int  d\phi_{\rm LO} b_2(s_1,z)  \right.
\nonumber \\
&&~~~+\int_z^1 dy\, y^{1-2\epsilon} \left[\frac{1}{(1-y)_+}-2\epsilon\left( \frac{{\rm ln}(1-y)}{1-y}\right)_+\right] \nonumber \\
&&~~~\left.  \times \int Nd\phi_{\rm LO}(p_1,p_2)  b_2(s_1,z)\right\}+{\cal O}(\epsilon),
\end{eqnarray}
and
\begin{eqnarray}
&&N_{CS}\int d\phi_{\rm real}\frac{c_4(s_1,z,y)}{s_2^2}\left[p_2\cdot p_3 +\frac{(y-z)M^2}{z}\left(\frac{r_b}{2}+\frac{1-y}{z}\right) \right. \nonumber \\
&&~~~\left. -\frac{1-y}{2z}(s_1-(1+r_c^2)M^2)\right]\nonumber \\
&&=\frac{\Gamma(1+\epsilon)(m_b^2)^{-\epsilon}}{\epsilon(4\pi)^{2-\epsilon}}\int_z^1 dy [(1-y)(1-y+r_b z)]^{-\epsilon}\left(\frac{y}{r_b z} \right)^{1-2\epsilon}\nonumber \\
&&~~~ \times \frac{y-z}{2r_bz}\int Nd\phi_{\rm LO}(p_1,p_2) c_4(s_1,z,y).
\end{eqnarray}

For the subtraction terms involving $s_3$, the parametrization in Eq.(\ref{eqb16}) is adopted and the expression for $d\phi^{(3)}(p_1,p_2,p_3)$ in the form of Eq.(\ref{phsp1}) is
\begin{eqnarray}
&&d\phi^{(3)}(p_1,p_2,p_3)\nonumber \\
&&=\frac{1}{4(2\pi)^{3-2\epsilon}}(y-r_b z)^{-1+\epsilon}y^{1-2\epsilon}(1-y)^{-\epsilon}\nonumber \\
&&~~~ \times\left(s_3-\frac{r_c(1-r_bz)(s_1-m_b^2)}{y-r_bz}\right)^{-\epsilon}dy\, ds_3\, d\Omega_{3\perp}.\label{phsp6}
\end{eqnarray}
The ranges of $y$ and $s_1$ are the same as above. The range of $s_3$ is from $r_c(1-r_bz)(s_1-m_b^2)/(y-r_bz)$ to $\infty$. Then we obtain
\begin{eqnarray}
&&N_{CS}\int d\phi_{\rm real}\frac{c_5(s_1,z,y)}{s_3}\nonumber \\
&&=\frac{\Gamma(1+\epsilon)[r_c(1-r_bz)]^{-\epsilon}}{\epsilon(4\pi)^{2-\epsilon}}\int_z^1 dy (1-y)^{-\epsilon}\left(\frac{y}{y-r_b z} \right)^{1-2\epsilon}\nonumber \\
&&~~~ \times\int Nd\phi_{\rm LO}(p_1,p_2) c_5(s_1,z,y)(s_1-m_b^2)^{-\epsilon},\\
&&N_{CS}\int d\phi_{\rm real}\frac{b_3(s_1,z)}{(1-y)s_3}\nonumber \\
&&=\frac{\Gamma(1+\epsilon)[r_c(1-r_bz)]^{-\epsilon}}{\epsilon(4\pi)^{2-\epsilon}}\left\{-\frac{1}{\epsilon} N_{CS}\int  d\phi_{\rm LO}\right. \nonumber \\
&& ~~~ \times b_3(s_1,z)(s_1-m_b^2)^{-\epsilon} (1-r_bz)^{-1+2\epsilon}
\nonumber \\
&&~~~ +\int_z^1 dy \left(\frac{y}{y-r_b z} \right)^{1-2\epsilon} \left[\frac{1}{(1-y)_+}-\epsilon\left( \frac{{\rm ln}(1-y)}{1-y}\right)_+\right] \nonumber \\
&& ~~~\left. \times \int Nd\phi_{\rm LO}(p_1,p_2)  b_3(s_1,z)(s_1-m_b^2)^{-\epsilon}\right\}+{\cal O}(\epsilon),
\end{eqnarray}
and
\begin{eqnarray}
&&N_{CS}\int d\phi_{\rm real}\frac{c_6(s_1,z,y)}{s_3^2}\nonumber \\
&& \times \left[p_1\cdot p_3+\frac{r_c z(1-r_b z)-(1-y)(y-z)}{2(y-r_b z)^2}(s_1-m_b^2) \right]\nonumber \\
&&=\frac{\Gamma(1+\epsilon)[r_c(1-r_bz)]^{-\epsilon}}{\epsilon(4\pi)^{2-\epsilon}}\int_z^1 dy (1-y)^{-\epsilon}\left(\frac{y}{y-r_b z} \right)^{1-2\epsilon}\nonumber \\
&& ~~~ \times\frac{z}{2(y-r_bz)}\int Nd\phi_{\rm LO}(p_1,p_2) c_6(s_1,z,y)(s_1-m_b^2)^{-\epsilon}.\nonumber \\
\end{eqnarray}

To integrate the subtraction terms that contain $t_1$, we adopt the parametrization in Eq.(\ref{eqb23}) for the differential phase space. The expression of the differential phase space in Eq.(\ref{eqb23}) can be written as
\begin{eqnarray}
N_{CS}d\phi_{\rm real}=N_{\rm LO}(p_1,\tilde{p})d\phi_{\rm LO}(p_1,\tilde{p})d\tilde{\phi}^{(3)}(p_1,p_2,p_3),\nonumber \\ \label{phsp7}
\end{eqnarray}
where the prefactor $N_{\rm LO}(p_1,\tilde{p})$ is defined as
\begin{eqnarray}
N_{\rm LO}(p_1,\tilde{p})=\frac{z^{1-2\epsilon}}{8\pi N_c},\label{phsp8}
\end{eqnarray}
and $d\phi_{\rm LO}(p_1,\tilde{p})$ is defined as
\begin{eqnarray}
d\phi_{\rm LO}(p_1,\tilde{p})=&&\frac{z^{-1+\epsilon}(1-z)^{-\epsilon}}{2(4\pi)^{1-\epsilon}\Gamma(1-\epsilon)K\cdot n} \nonumber \\
&&\times \left(\tilde{s}-\frac{M^2}{z}-\frac{m_c^2}{1-z}\right)^{-\epsilon}d\tilde{s}.\label{phsp9}
\end{eqnarray}
Then the expression of $d\tilde{\phi}^{(3)}(p_1,p_2,p_3)$ can be written as
\begin{eqnarray}
d\tilde{\phi}^{(3)}(p_1,p_2,p_3)=&&\frac{(1/z-1)^{1-\epsilon}}{4(2\pi)^{3-2\epsilon}}\frac{u^{-\epsilon}}{1-u}\left[t_1-\frac{(1-z)M^2 u}{z}\right]^{-\epsilon}\nonumber \\
&&\times du\, dt_1 d\Omega_{3\perp},\label{phsp10}
\end{eqnarray}
The range of $\tilde{s}$ is from $[M^2/z+ m_c^2/(1-z)]$ to $\infty$, the range of $u$ is from 0 to 1, and the range of $t_1$ is from $(1/z-1)M^2u$ to $\infty$.

After integrating over $\Omega_{3\perp}$, $t_1$ and $u$, we obtain
\begin{eqnarray}
&&N_{CS}\int d\phi_{\rm real}\frac{d_1(\tilde{s},z)(1-u)(\tilde{s}-m_b^2)}{u~t_1(\tilde{s}-m_b^2+t_1/z)}\nonumber \\
&&=\frac{\Gamma(1+\epsilon)}{\epsilon(4\pi)^{2-\epsilon}}\left(\frac{1-z}{z}\right)^{1-\epsilon} \int Nd\phi_{\rm LO}(p_1,\tilde{p}) d_1(\tilde{s},z)\nonumber \\
&&~~~ \times\left\{ -\frac{1}{2\epsilon}\left[\frac{(1-z)M^2}{z}\right]^{-\epsilon} + \frac{1}{\epsilon}[z(\tilde{s}-m_b^2)]^{-\epsilon} \right.\nonumber \\
&&~~~ \left. -\epsilon \, {\rm Li}_2\left[-\frac{(1-z)M^2}{z^2(\tilde{s}-m_b^2)} \right]\right\}+{\cal O}(\epsilon),
\end{eqnarray}
\begin{eqnarray}
&&N_{CS}\int d\phi_{\rm real}\frac{d_2(\tilde{s},z)(1-u)(\tilde{s}-m_b^2)}{u~t_1(\tilde{s}-m_b^2+(1-r_bz)t_1/(r_cz))}\nonumber \\
&&=\frac{\Gamma(1+\epsilon)}{\epsilon(4\pi)^{2-\epsilon}}\left(\frac{1-z}{z}\right)^{1-\epsilon} \int Nd\phi_{\rm LO}(p_1,\tilde{p}) d_2(\tilde{s},z)\nonumber \\
&&~~~ \times\left\{ -\frac{1}{2\epsilon}\left[\frac{(1-z)M^2}{z}\right]^{-\epsilon} + \frac{1}{\epsilon}\left[\frac{r_c z}{1-r_bz}(\tilde{s}-m_b^2)\right]^{-\epsilon} \right.\nonumber \\
&&~~~ \left. -\epsilon \, {\rm Li}_2\left[-\frac{(1-z)(1-r_bz)M^2}{r_c z^2(\tilde{s}-m_b^2)} \right]\right\}+{\cal O}(\epsilon),
\end{eqnarray}
and
\begin{eqnarray}
&&N_{CS}\int d\phi_{\rm real}\frac{d_3(\tilde{s},z)(1-u)(\tilde{s}-m_b^2)^2}{u~t_1\left(\tilde{s}-m_b^2+\frac{t_1}{z}\right)\left(\tilde{s}-m_b^2+\frac{(1-r_b z)t_1}{r_c z}\right)}\nonumber \\
&&=\frac{\Gamma(1+\epsilon)}{\epsilon(4\pi)^{2-\epsilon}}\left(\frac{1-z}{z}\right)^{1-\epsilon} \int Nd\phi_{\rm LO}(p_1,\tilde{p}) d_3(\tilde{s},z)\nonumber \\
&&~~~ \times\left\{ -\frac{1}{2\epsilon}\left[\frac{(1-z)M^2}{z}\right]^{-\epsilon} - \frac{1}{\epsilon}\frac{r_c}{r_b(1-z)}\right. \nonumber \\
&&~~~ \times[z(\tilde{s}-m_b^2)]^{-\epsilon}\left[1-\left(\frac{r_c}{1-r_bz}\right)^{-1-\epsilon}\right] \nonumber \\
&&~~~  +\frac{\epsilon}{r_b(1-z)}\left[r_c\, {\rm Li}_2\left(-\frac{(1-z)M^2}{z^2(\tilde{s}-m_b^2)} \right)\right.\nonumber \\
&&~~~ \left. \left.-(1-r_bz) {\rm Li}_2\left(-\frac{(1-z)(1-r_bz)M^2}{r_c z^2(\tilde{s}-m_b^2)} \right) \right] \right\}+{\cal O}(\epsilon).\nonumber \\
\end{eqnarray}

For the subtraction terms that contain $t_2$, the parametrization in Eq.(\ref{eqb26}) is adopted and the expression for $d\tilde{\phi}^{(3)}(p_1,p_2,p_3)$ in the form of Eq.(\ref{phsp7}) is
\begin{eqnarray}
d\tilde{\phi}^{(3)}(p_1,p_2,p_3)=&&\frac{1}{4(2\pi)^{3-2\epsilon}}u^{-\epsilon}\left[(1-u)t_2-m_c^2 u\right]^{-\epsilon}\nonumber \\
&&\times du\, dt_2 d\Omega_{3\perp}.\label{phsp11}
\end{eqnarray}
The ranges of $\tilde{s}$ and $u$ are the same as above. The range of $t_2$ is from $u\,m_c^2/(1-u)$ to $\infty$. Then we obtain
\begin{eqnarray}
&&N_{CS}\int d\phi_{\rm real}\frac{d_4(\tilde{s},z)(\tilde{s}-m_b^2)^2}{u~t_2\left(\tilde{s}-m_b^2+\frac{t_2}{1-z}\right)\left(\tilde{s}-m_b^2+\frac{(1-r_b z)t_2}{r_c (1-z)}\right)}\nonumber \\
&&=\frac{\Gamma(1+\epsilon)}{\epsilon(4\pi)^{2-\epsilon}} \int Nd\phi_{\rm LO}(p_1,\tilde{p}) d_4(\tilde{s},z)\left\{ -\frac{1}{2\epsilon}\left(m_c^2\right)^{-\epsilon}\right.\nonumber \\
&&~~~ - \frac{1}{\epsilon}\frac{r_c}{r_b(1-z)}[(1-z)(\tilde{s}-m_b^2)]^{-\epsilon} \left[1-\left(\frac{r_c}{1-r_bz}\right)^{-1-\epsilon}\right] \nonumber \\
&&~~~  +\frac{\epsilon}{r_b(1-z)}\left[r_c\, {\rm Li}_2\left(1-\frac{m_c^2}{(1-z)(\tilde{s}-m_b^2)} \right)\right.\nonumber \\
&&~~~ \left. \left.-(1-r_bz) {\rm Li}_2\left(1-\frac{(1-r_bz)m_c^2}{r_c (1-z)(\tilde{s}-m_b^2)} \right) \right] \right\}+{\cal O}(\epsilon),
\end{eqnarray}
\begin{eqnarray}
&&N_{CS}\int d\phi_{\rm real}\frac{d_5(\tilde{s},z)(\tilde{s}-m_b^2)}{u~t_2(\tilde{s}-m_b^2+t_2/(1-z))}\nonumber \\
&&=\frac{\Gamma(1+\epsilon)}{\epsilon(4\pi)^{2-\epsilon}}\int Nd\phi_{\rm LO}(p_1,\tilde{p}) d_5(\tilde{s},z)\nonumber \\
&&~~~ \times\left\{ -\frac{1}{2\epsilon}\left(m_c^2\right)^{-\epsilon} + \frac{1}{\epsilon}[(1-z)(\tilde{s}-m_b^2)]^{-\epsilon} \right.\nonumber \\
&&~~~ \left. -\epsilon \, {\rm Li}_2\left[1-\frac{m_c^2}{(1-z)(\tilde{s}-m_b^2)} \right]\right\}+{\cal O}(\epsilon),
\end{eqnarray}
\begin{eqnarray}
&&N_{CS}\int d\phi_{\rm real}\frac{d_6(\tilde{s},z)(\tilde{s}-m_b^2)}{u~t_2[\tilde{s}-m_b^2+(1-r_bz)t_2/(r_c(1-z))]}\nonumber \\
&&=\frac{\Gamma(1+\epsilon)}{\epsilon(4\pi)^{2-\epsilon}}\int Nd\phi_{\rm LO}(p_1,\tilde{p}) d_6(\tilde{s},z)\nonumber \\
&&~~~ \times\left\{ -\frac{1}{2\epsilon}\left(m_c^2\right)^{-\epsilon} + \frac{1}{\epsilon}\left[\frac{r_c(1-z)(\tilde{s}-m_b^2)}{1-r_bz}\right]^{-\epsilon} \right.\nonumber \\
&&~~~ \left. -\epsilon \, {\rm Li}_2\left[1-\frac{(1-r_bz)m_c^2}{r_c(1-z)(\tilde{s}-m_b^2)} \right]\right\}+{\cal O}(\epsilon),
\end{eqnarray}
\begin{eqnarray}
&&N_{CS}\int d\phi_{\rm real}\frac{g(\tilde{s},z)(\tilde{s}-m_b^2)^2}{u\left(\tilde{s}-m_b^2+\frac{t_2}{1-z}\right)\left(\tilde{s}-m_b^2+\frac{(1-r_b z)t_2}{r_c (1-z)}\right)}\nonumber \\
&&=\frac{\Gamma(1+\epsilon)r_c}{\epsilon(4\pi)^{2-\epsilon}r_b} \int Nd\phi_{\rm LO}(p_1,\tilde{p}) g(\tilde{s},z)(\tilde{s}-m_b^2)\nonumber \\
&&~~~ \left\{- \frac{1}{\epsilon}[(1-z)(\tilde{s}-m_b^2)]^{-\epsilon} \left[\left(\frac{r_c}{1-r_bz}\right)^{-\epsilon}-1\right]\right. \nonumber \\
&&~~~  +\epsilon \left[-{\rm Li}_2\left(1-\frac{m_c^2}{(1-z)(\tilde{s}-m_b^2)} \right)\right.\nonumber \\
&&~~~ \left. \left.+ {\rm Li}_2\left(1-\frac{(1-r_bz)m_c^2}{r_c (1-z)(\tilde{s}-m_b^2)} \right) \right] \right\}+{\cal O}(\epsilon),
\end{eqnarray}
and
\begin{eqnarray}
&&N_{CS}\int d\phi_{\rm real}\frac{h(\tilde{s},z)}{t_2^2}\nonumber \\
&&=-\frac{\Gamma(1+\epsilon)}{(4\pi)^{2-\epsilon}}\frac{(m_c^2)^{-1-\epsilon}}{2\epsilon(1-2\epsilon)} \int Nd\phi_{\rm LO}(p_1,\tilde{p}) h(\tilde{s},z).
\end{eqnarray}

Since the remaining integrals in these expressions do not generate poles in $\epsilon$, we can expand these expressions in powers of $\epsilon$ before performing the integration.

\subsection{The renormalization}

There are UV divergences remaining after summing the contributions from virtual and real corrections, while they are removed by renormalization. We adopt the counterterm approach to carry out the renormalization, where the FFs are calculated with the renormalized coupling constant $g_s$, the renormalized quark mass $m$, field $\Psi_r$\footnote{Here the mass $m$ and field $\Psi_r$ may be the mass and field of a $b$ quark or $c$ quark.}, and the renormalized gluon field $A^{\mu}_r$. The renormalized quantities are related to their corresponding bare quantities as
\begin{eqnarray}
&&g_s^0= Z_g \,g_s, ~~~~~m^0=Z_m \,m, \nonumber \\
&&\Psi_0=\sqrt{Z_2} \,\Psi_r, \, A^{\mu}_0=\sqrt{Z_3} \,A^{\mu}_r,
\end{eqnarray}
where $Z_i = 1 + \delta Z_i$ with $i=g, m, 2, 3$ are renormalization constants. The quantities $\delta Z_i$ are fixed by the precise definitions of the renormalized quantities. The renormalized quark field, quark mass, and gluon field are defined in the on-mass-shell scheme (OS), whereas the renormalized strong coupling constant $g_s$ is defined in the modified-minimal-subtraction
scheme ($\overline{\rm MS}$). The expressions of the corresponding renormalization constants in this scheme are obtained as follows:
\begin{eqnarray}
\label{rencont}
\delta Z^{OS}_2&=&-C_F \frac{\alpha_s(\mu_R)}{4\pi}\left[\frac{1}{\epsilon_{UV}}+ \frac{2}{\epsilon_{IR}}-3~\gamma_E+3~ {\rm ln}\frac{4\pi \mu_R^2}{m^2}+4\right], \nonumber\\
\delta Z^{OS}_m&=&-3~C_F \frac{\alpha_s(\mu_R)}{4\pi}\left[\frac{1}{\epsilon_{UV}}- \gamma_E+
 {\rm ln}\frac{4\pi \mu_R^2}{m^2}+\frac{4}{3}\right],\nonumber\\
 \delta Z^{OS}_3&=&\frac{\alpha_s(\mu_R)}{4\pi}\left[(\beta'_0-2C_A)\left(\frac{1}{\epsilon_{UV}}-\frac{1}{\epsilon_{IR}}\right) \right. \nonumber\\
 &&\left.-\frac{4}{3}T_F \left(\frac{1}{\epsilon_{UV}}-\gamma_E + {\rm ln}\frac{4\pi \mu_R^2}{m_c^2}\right)\right. \nonumber\\
 &&\left.-\frac{4}{3}T_F \left(\frac{1}{\epsilon_{UV}}-\gamma_E + {\rm ln}\frac{4\pi \mu_R^2}{m_b^2}\right)\right], \nonumber\\
 \delta Z^{\overline{MS}}_g&=&- \frac{\beta_0}{2}\frac{\alpha_s(\mu_R)}{4\pi}\left[\frac{1}{\epsilon_{UV}}- \gamma_E+ {\rm ln}~(4\pi)\right],
\end{eqnarray}
where $\mu_R$ is the renormalization scale, $\beta_0=\frac{11}{3}C_A-\frac{4}{3} T_F n_f$ is the one-loop coefficient of the $\beta$ function in QCD, $n_f$ is the number of active quark flavors, $\beta'_0=\frac{11}{3}C_A-\frac{4}{3} T_F n_{lf}$ and $n_{lf} = 3$ is the number of the light-quark flavors.

Then the contribution from these counterterms can be expressed as
\begin{eqnarray}
\label{eqcts}
D^{\rm counter}_{\bar{b}\to c\bar{b}[n]}(z)=N_{CS}\int d\phi_{\rm LO} {\cal A}_{\rm counter},
\label{eqcounter}
\end{eqnarray}
where ${\cal A}_{\rm counter}$ denotes the squared amplitudes for the counterterms from the renormalization of the quark field, the gluon field, the quark mass and the strong coupling.

Obviously the NLO FFs defined as in Ref.\cite{Collins} by operator products require renormalization\cite{Mueller}. We carry out the operator renormalization in the $\overline{\rm MS}$ scheme. The expression for the counterterms for the operator
products in this scheme is
\begin{eqnarray}\label{eqmuF}
&&D^{\rm operator}_{\bar{b}\to c\bar{b}[n]}(z)\nonumber \\
&&=-\frac{\alpha_s(\mu_R)}{2\pi}\left[\frac{1}{\epsilon_{UV}}- \gamma_E+ {\rm ln}~(4\pi)+{\rm ln}\frac{\mu_R^2}{\mu_F^2} \right]\nonumber \\
&&\times \int_z^1 \frac{dy}{y}P_{\bar{b}\bar{b}}(y)D_{\bar{b}\to c\bar{b}[n]}^{\rm LO}(z/y), \;\;\;
\end{eqnarray}
where $\mu_F$ is the factorization scale for the FFs and $D_{\bar{b}\to c\bar{b}[n]}^{\rm LO}(z)$ denotes the LO FFs in $d$-dimensional space-time.

\subsection{The numerical results}

Canceling the pole terms in $\epsilon$, the NLO FFs can be obtained by summing the finite parts from virtual and real corrections and counterterms:
\begin{eqnarray}\label{FF0-NLO}
&&D_{\bar{b} \to c\bar{b}[n]}^{NLO}(z,\mu_F,\mu_R)=D_{\bar{b} \to c\bar{b}[n]}^{LO}(z,\mu_R)\nonumber\\
&&~~ +D_{\bar{b} \to c\bar{b}[n]}^{\rm virtual}(z,\mu_R)+D_{\bar{b} \to c\bar{b}[n]}^{\rm real}(z,\mu_R) \nonumber\\
&&~~ +D_{\bar{b} \to c\bar{b}[n]}^{\rm counter}(z,\mu_R) +D_{\bar{b} \to c\bar{b}[n]}^{\rm operator}(z,\mu_F,\mu_R),
\end{eqnarray}
where the terms on the right-hand side of the equation are defined in Eqs.(\ref{eqlo}),(\ref{eqvir}),(\ref{eqreal}),(\ref{dsub}),(\ref{eqcounter}) and (\ref{eqmuF}), and the renormalization and factorization scales are written explicitly here. The FFs $D_{\bar{b}\to B_c(B_c^*)}^{NLO}(z,\mu_F,\mu_R)$ can be obtained by multiplying the matrix element $\langle {\cal O} ^{B_c(B_c^*)}(n)\rangle/ \langle {\cal O} ^{c\bar{b}[n]}(n)\rangle \approx \vert R_S(0) \vert^2/4\pi$ to $D_{\bar{b} \to c\bar{b}[n]}^{NLO}(z,\mu_F,\mu_R)$, where $n=\, ^1S_0$ or $^3S_1$ accordingly. In the numerical calculations, the integrations over phase space are performed numerically with the help of the program Vegas\cite{vegas}.

The necessary input masses in the numerical calculations are taken as follows:
\begin{eqnarray}
&& m_b=4.9 ~{\rm GeV}\,,\; m_c=1.5 ~{\rm GeV}\,,\nonumber \\
&&m_{_Z}=91.1876~{\rm GeV}.
\end{eqnarray}
The value of $\vert R_S(0) \vert^2$ may be extracted from the experimental widths of the $B_c$ pure leptonic decays, potential-model calculations and lattice QCD calculations, etc.,
whereas now there is no very accurate value of $\vert R_S(0) \vert^2$. In fact, due to the fact that for the FFs $\vert R_S(0) \vert^2$ is an overall factor, the numerical results obtained in this paper with a given value of $\vert R_S(0) \vert^2$ can be easily updated with a more accurate value. Thus, as an approximation, in numerical calculations we just take the value from the potential-model calculations\cite{pot}:
\begin{eqnarray}
\vert R_S(0) \vert^2=1.64 ~{\rm GeV^3}.
\end{eqnarray}
For strong coupling constant, we adopt the two-loop formula
\begin{equation}
\alpha_s(\mu)=\frac{4\pi}{\beta_0~{\rm ln}(\mu^2/\Lambda^2_{QCD})}\left[ 1-\frac{\beta_1~{\rm ln}~{\rm ln}(\mu^2/\Lambda^2_{QCD})}{\beta_0^2~{\rm ln}(\mu^2/\Lambda^2_{QCD})}\right],
\end{equation}
where $\beta_1=\frac{34}{3}C_A^2-4C_F T_F n_f-\frac{20}{3}C_A T_F n_f$ is the two-loop coefficient of $\beta$ function in QCD. According to $\alpha_s(m_{_Z})=0.1185$\cite{pdg}, we obtain $\Lambda_{QCD}^{n_f=5}=0.233 {\rm GeV}$ and $\Lambda_{QCD}^{n_f=4}=0.337 {\rm GeV}$. Then we have $\alpha_s(2m_c)=0.259$, $\alpha_s(m_b+2m_c)=0.190$, $\alpha_s(2m_b)=0.180$, and $\alpha_s(2m_b+m_c)=0.174$.

\begin{figure}[htbp]
\includegraphics[width=0.5\textwidth]{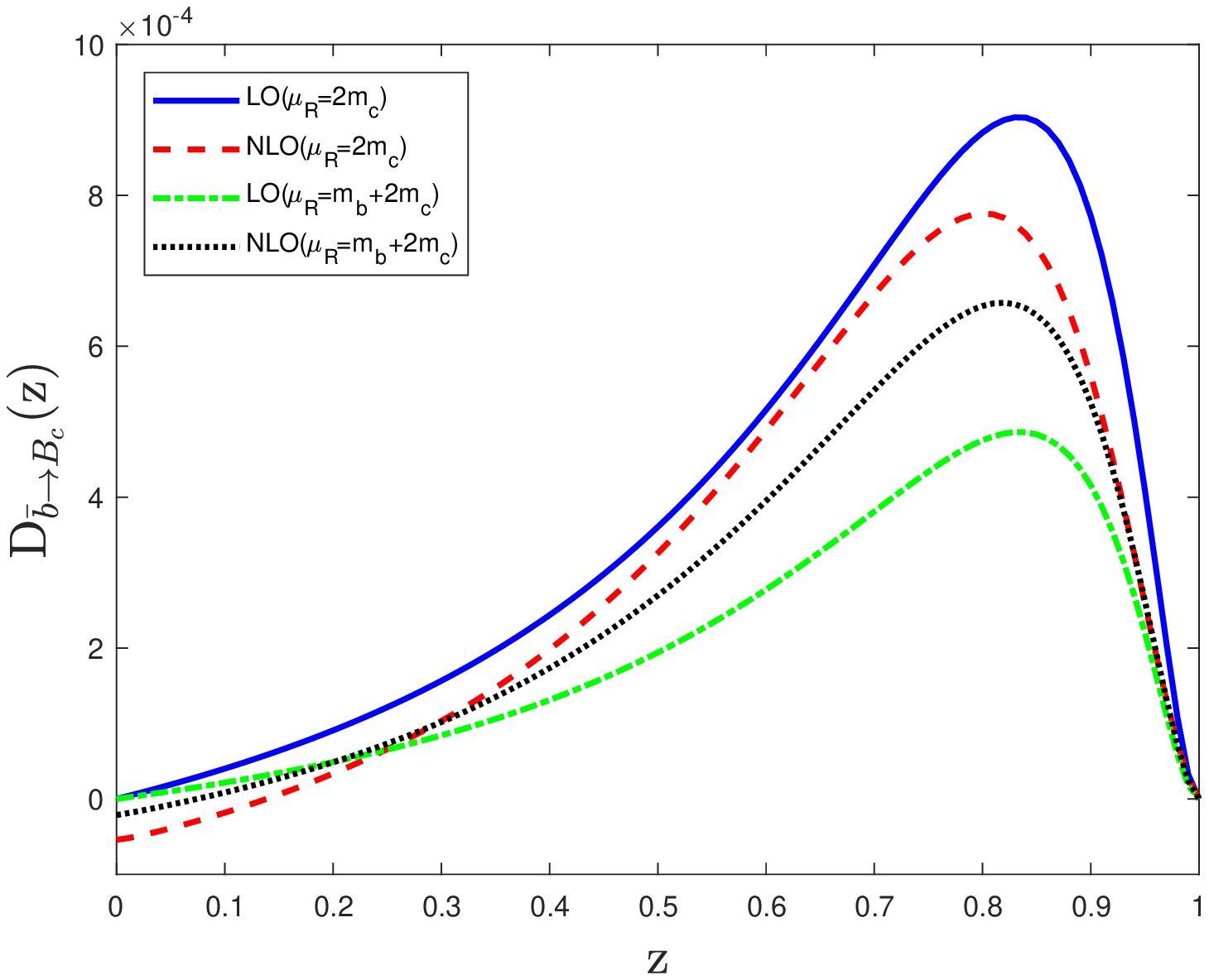}
\caption{The initial FF $D_{\bar{b}\to B_c}(z,\mu_{F0},\mu_R)$ as a function of $z$ with $\mu_{F0}=m_b+2m_c$, $\mu_R=2m_c$, or $\mu_R=m_b+2m_c$ up to LO and NLO accuracy.} \label{dz1s0}
\end{figure}

\begin{figure}[htbp]
\includegraphics[width=0.5\textwidth]{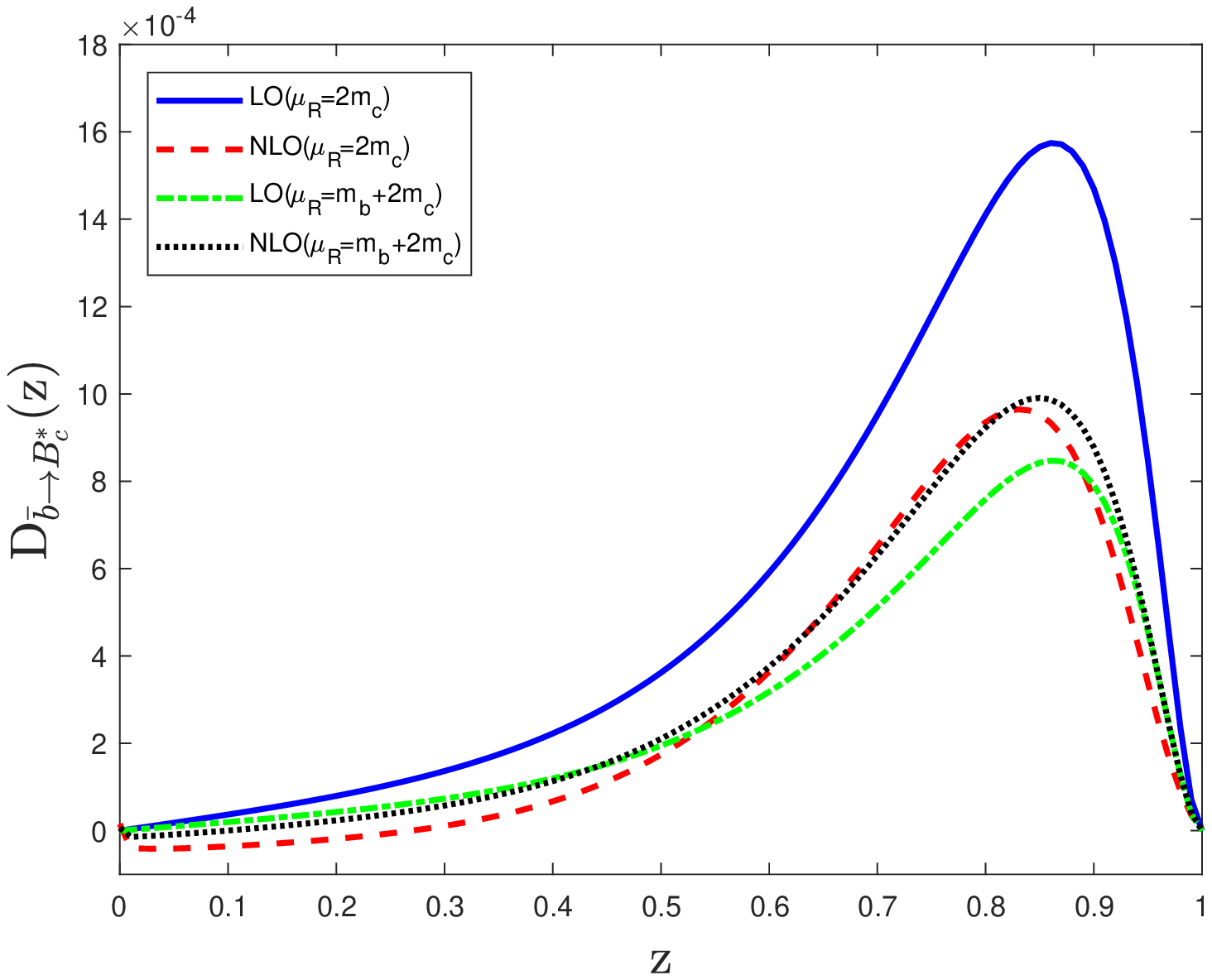}
\caption{The initial FF $D_{\bar{b}\to B^*_c}(z,\mu_{F0},\mu_R)$ as a function of $z$ with $\mu_{F0}=m_b+2m_c$, $\mu_R=2m_c$, or $\mu_R=m_b+2m_c$ up to LO and NLO accuracy.} \label{dz3s1}
\end{figure}

The LO FFs $D^{LO}_{\bar{b}\to B_c}(z,\mu_{F0},\mu_R)$, $D^{LO}_{\bar{b}\to B^*_c}(z,\mu_{F0},\mu_R)$ and the NLO FFs $D^{NLO}_{\bar{b}\to B_c}(z,\mu_{F0},\mu_R)$, $D^{NLO}_{\bar{b}\to B^*_c}(z,\mu_{F0},\mu_R)$(the latter is that in Eq.(\ref{FF0-NLO}) are presented in Figs.\ref{dz1s0} and \ref{dz3s1}, respectively. In order to keep the logarithm terms ${\rm ln}(\mu_R/m_Q)$ and ${\rm ln}(\mu_{F0}/m_Q)$ ($m_Q, Q=c,b$) in higher-order corrections ``becoming" large and to have better accuracy, here we set the renormalization scale $\mu_R$ and factorization scales $\mu_{F0}$ to ${\cal O}(m_Q)$, i.e., we set $\mu_R$ and $\mu_{F0}$ to be $2m_c$ and $m_b+2m_c$(the minimum invariant mass of the initial off-shell $\bar{b}$ quark), respectively. In Figs.\ref{dz1s0} and \ref{dz3s1} the results for $\mu_R=\mu_{F0}=m_b+2m_c$ are also presented.

From Figs.\ref{dz1s0} and \ref{dz3s1}, one can see that the QCD NLO corrections to the FFs of $\bar{b}$ quark are quite large with a normalization scale $\mu_R=2m_c$ or $\mu_R=m_b+2m_c$. The maximum points of the FFs are shifted to smaller values of $z$ when the NLO corrections are involved. Moreover, the QCD NLO FFs are scheme and scale dependent, and the FFs in this paper are defined in the $\overline{MS}$ scheme.

There are two useful quantities which can be easily computed from the numerical results for the FFs: the fragmentation probability $P$ and the average value of $z$, $\langle z \rangle$. They are defined as follows:
\begin{eqnarray}
P=\int_0^1 dz D(z),~~ \langle z \rangle =\frac{\int_0^1 dz\,z\, D(z)}{\int_0^1 dz\, D(z)},
\end{eqnarray}
where $D(z)$ denotes an FF at a given energy scale. The numerical results for the obtained FFs are presented in Tables \ref{tbcfp} and \ref{tbc*fp}. From the two tables, one can see that the NLO corrections to the fragmentation probabilities are sizable with the two choices of the renormalization scale. However, due to the QCD NLO corrections the average values $\langle z \rangle$ change by only a small amount.
\begin{table}[h]
\begin{tabular}{c c c c c}
\hline\hline
$\mu_R$ &  ${\rm P}\times 10^{4}$(LO)&  ${\rm P}\times 10^{4}$(NLO) &$\langle z \rangle$(LO) &  $\langle z \rangle$(NLO)\\
\hline
$2 m_c$ &  3.82 & 3.14  & 0.68 & 0.70  \\
$m_b+2 m_c$ & 2.05 & 2.73  & 0.68 & 0.69 \\
\hline\hline
\end{tabular}
\caption{The fragmentation probability and average value of $z$ for $D_{\bar{b}\to B_c}(z,\mu_{F0}=m_b+2m_c,\mu_R)$ with two typical renormalization scales.}
\label{tbcfp}
\end{table}

\begin{table}[h]
\begin{tabular}{c c c c c}
\hline\hline
$\mu_R$ &  ${\rm P}\times 10^{4}$(LO) &  ${\rm P}\times 10^{4}$(NLO) &$\langle z \rangle$(LO) &  $\langle z \rangle$(NLO)\\
\hline
$2 m_c$ &  5.36 & 2.91  & 0.73 &   0.77 \\
$m_b+2 m_c$ & 2.89 & 3.25  & 0.73 & 0.74 \\
\hline\hline
\end{tabular}
\caption{The fragmentation probability and average value of $z$ for $D_{\bar{b}\to B^*_c}(z,\mu_{F0}=m_b+2m_c,\mu_R)$ with two typical renormalization scales.}
\label{tbc*fp}
\end{table}

\begin{figure}[htbp]
\includegraphics[width=0.5\textwidth]{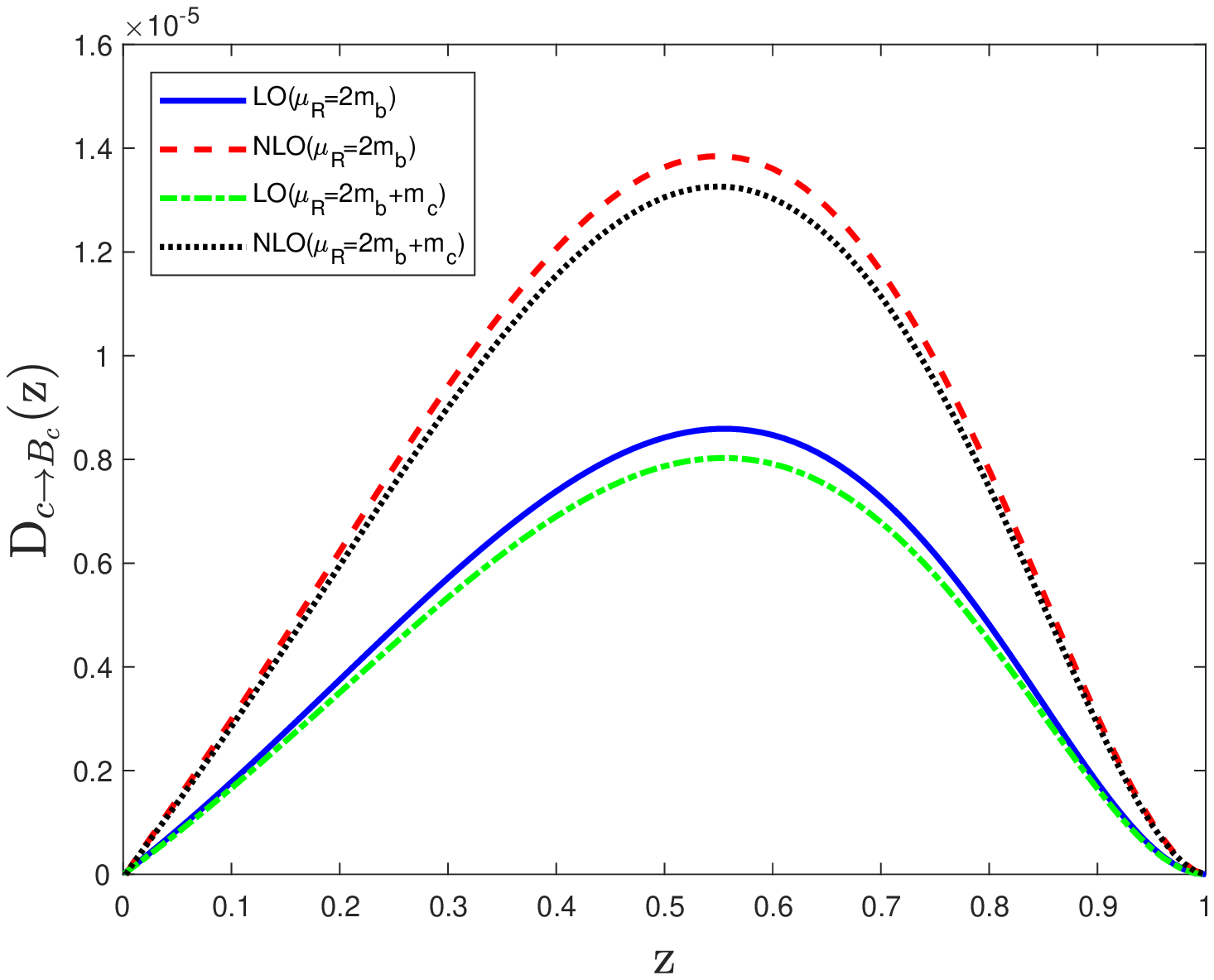}
\caption{The initial FF $D_{c \to B_c}(z,\mu_{F0}=2m_b+m_c,\mu_R)$ as a function of $z$ with two typical renormalization scales ($\mu_R=2m_b$ or $\mu_R=2m_b+m_c$) up to LO and NLO accuracy.} \label{dzc1s0}
\end{figure}

\begin{figure}[htbp]
\includegraphics[width=0.5\textwidth]{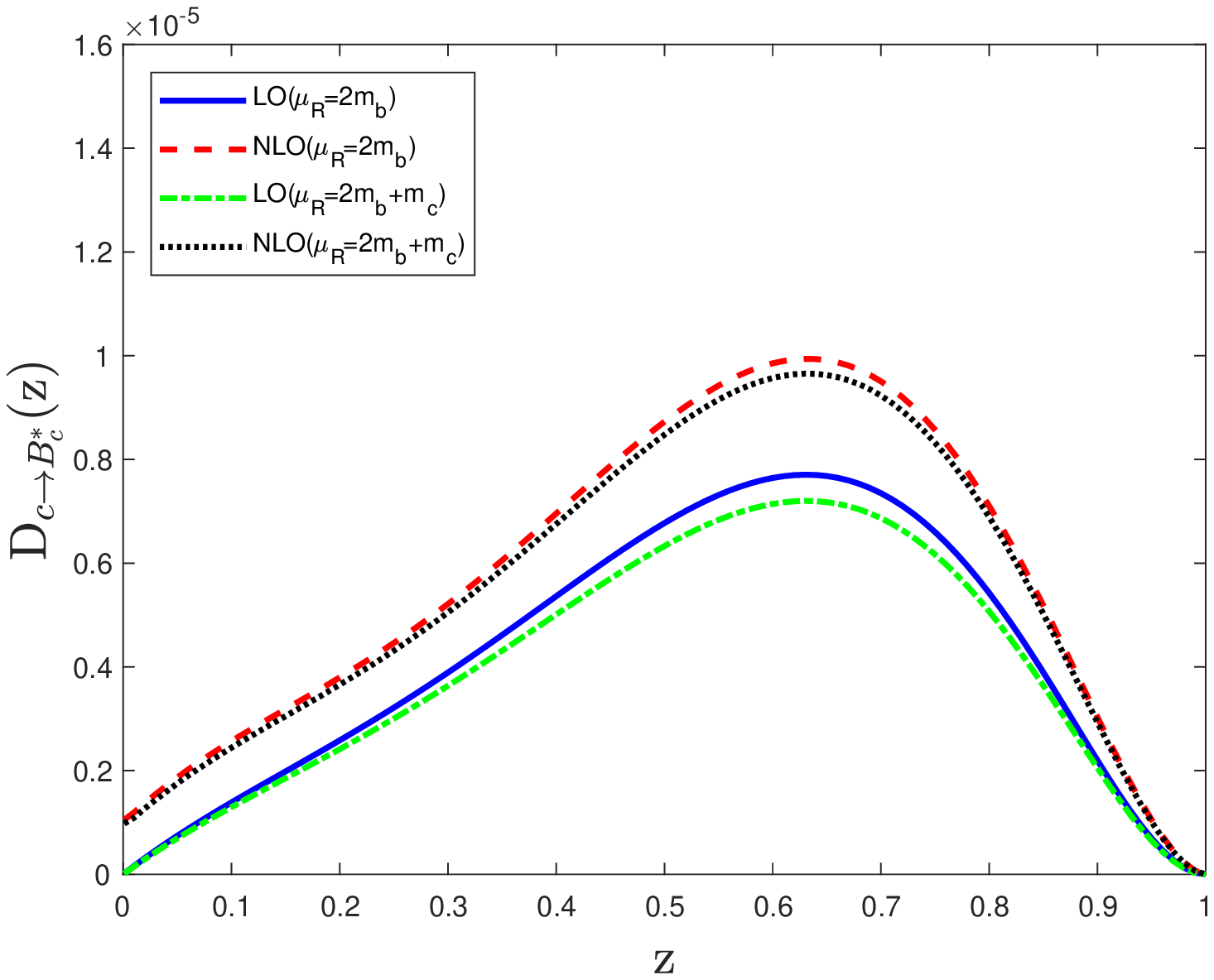}
\caption{The initial FF $D_{c\to B^*_c}(z,\mu_{F0}=2m_b+m_c,\mu_R)$ as a function of $z$ with two typical renormalization scales ($\mu_R=2m_b$ or $\mu_R=2m_b+m_c$) up to LO and NLO accuracy.} \label{dzc3s1}
\end{figure}

The FFs of a c quark to the meson $B_c$ or $B^*_c$ can be
derived out by applying the method presented in Secs. II
and III precisely. Whereas, the contributions to the FFs
from the cut diagrams without a heavy-quark loop on either
side of the cut can be obtained by the alternation of $m_b$ and $m_c$. The NLO QCD FFs of a $c$ quark into $B_c$ and $B^*_c$ mesons are presented in Figs.\ref{feyvir} and \ref{feyreal} with two possible renormalization scales, $\mu_R=2m_b$ and $\mu_R=2m_b+m_c$, and the factorization scale is set $\mu_{F0}=2m_b+m_c$, which is the minimum invariant mass of the initial off-shell $c$ quark.

From Fig.\ref{dzc1s0} and Fig.\ref{dzc3s1}, one can see that the NLO QCD corrections to the FFs of $B_c$ and $B^*_c$ are also sizable with the renormalization scales $\mu_R=2m_b$ or $\mu_R=2m_b+m_c$, and the difference between the FFs at the two renormalization scales is quite small. This is because these two scales are quite close to each other.
\begin{table}[h]
\begin{tabular}{c c c c c}
\hline\hline
$\mu_R$ &  ${\rm P}\times 10^{6}$(LO) &  ${\rm P}\times 10^{6}$(NLO) &$\langle z \rangle$(LO) &  $\langle z \rangle$(NLO)\\
\hline
$2 m_b$ &  4.95 & 8.07  & 0.51 & 0.51  \\
$2m_b+m_c$ & 4.63 & 7.72  & 0.51 & 0.51 \\
\hline\hline
\end{tabular}
\caption{The fragmentation probability and average value of $z$ for $D_{c\to B_c}(z,\mu_{F0}=2m_b+m_c,\mu_R)$ with two typical renormalization scales.}
\label{tcbcfp}
\end{table}

\begin{table}[h]
\begin{tabular}{c c c c c}
\hline\hline
$\mu_R$ ~&  ${\rm P}\times 10^{6}$(LO)&  ${\rm P}\times 10^{6}$(NLO) &$\langle z \rangle$(LO) &  $\langle z \rangle$(NLO)\\
\hline
$2 m_b$ &  4.28 & 5.75  & 0.55 &   0.54 \\
$2m_b+ m_c$ & 4.00 & 5.57  & 0.55 & 0.54 \\
\hline\hline
\end{tabular}
\caption{The fragmentation probability and average value of $z$ for $D_{c\to B^*_c}(z,\mu_{F0}=2m_b+m_c,\mu_R)$ with two typical renormalization scales.}
\label{tcbc*fp}
\end{table}

The fragmentation probabilities and average values of $z$ for $c$-quark fragmentation are presented in Tables \ref{tcbcfp} and \ref{tcbc*fp}. One can see that the fragmentation probability of $c \to B_c$($c \to B^*_c$) is smaller than that of $\bar{b} \to B_c$($\bar{b} \to B_c$) by about 2 orders of magnitude.

The FFs at a large factorization scale such as at $\mu_F \gg m_Q$ can be obtained by solving the DGLAP evolution equations from the FFs at a smaller $\mu_{F0}$ ($\sim m_Q$). Note that for convenience in the paper we call the FFs at a smaller factorization scale as ``initial FFs".

Here to solve the DGLAP evolution equations the approximation method introduced in Ref.\cite{APsolve} is adopted, and as stated in the Introduction, the evolution of FFs are from a low energy scale to a high energy scale is restricted to the LL QCD level, namely, only the LO splitting function $P_{ij}$ ($i,j=g,q$, where $g$ is gluon and the quarks $q=b,c$) in Eq.(\ref{eq.spfun1}) is considered.

\begin{figure}[htbp]
\includegraphics[width=0.5\textwidth]{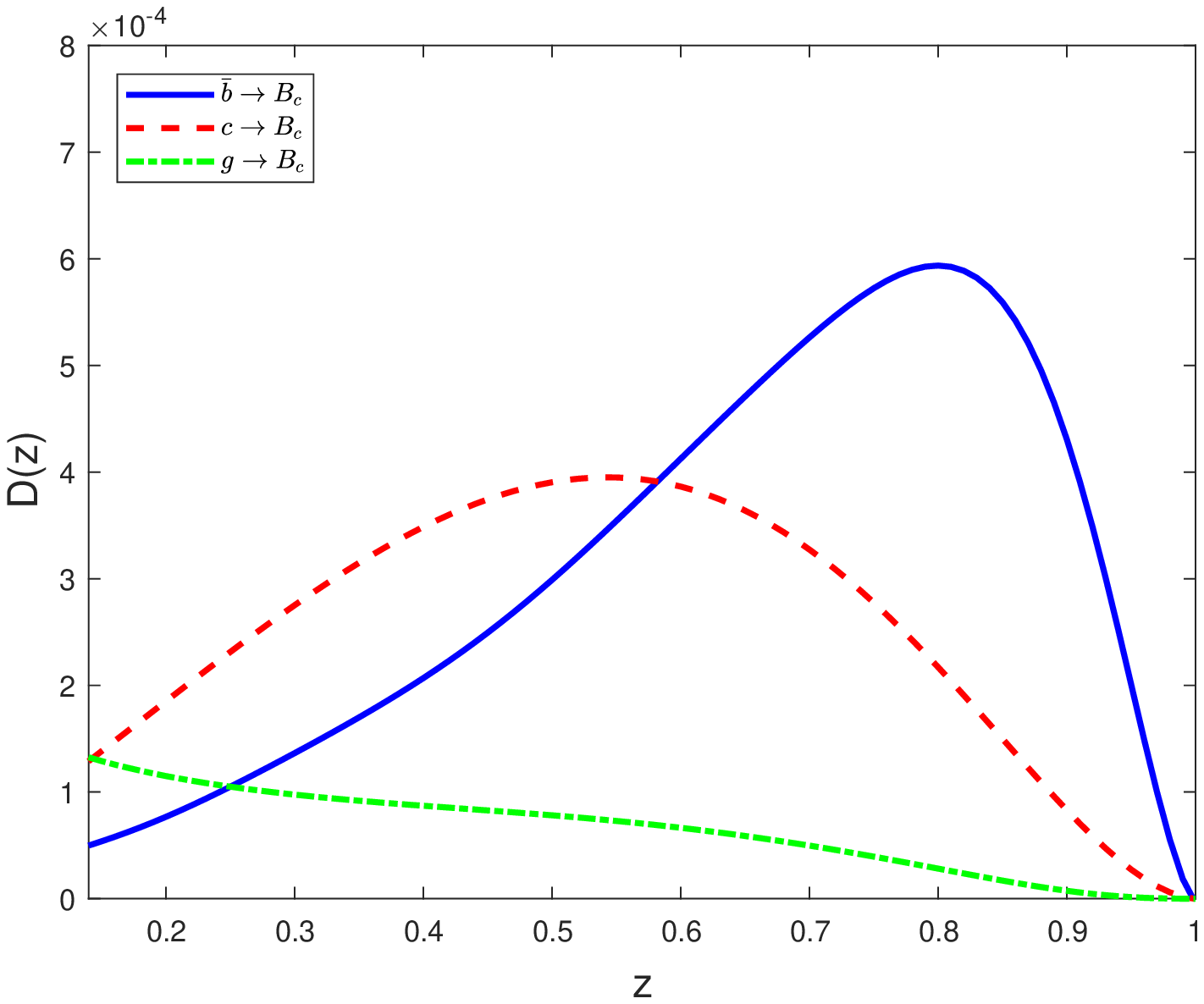}
\caption{The FFs $D_{\bar{b} \to B_c}(z,\mu_{F})$, $D_{c \to B_c}(z,\mu_{F})$ and $D_{g \to B_c}(z,\mu_{F})$ as functions of $z$ with $\mu_{F}=2m_b+2m_c$. In order to show these results in one figure, $D_{c \to B_c}(z,\mu_{F})$ and $D_{g \to B_c}(z,\mu_{F})$ are artificially multiplied by a factor of 30.} \label{dzmbc1s0}
\end{figure}

\begin{figure}[htbp]
\includegraphics[width=0.5\textwidth]{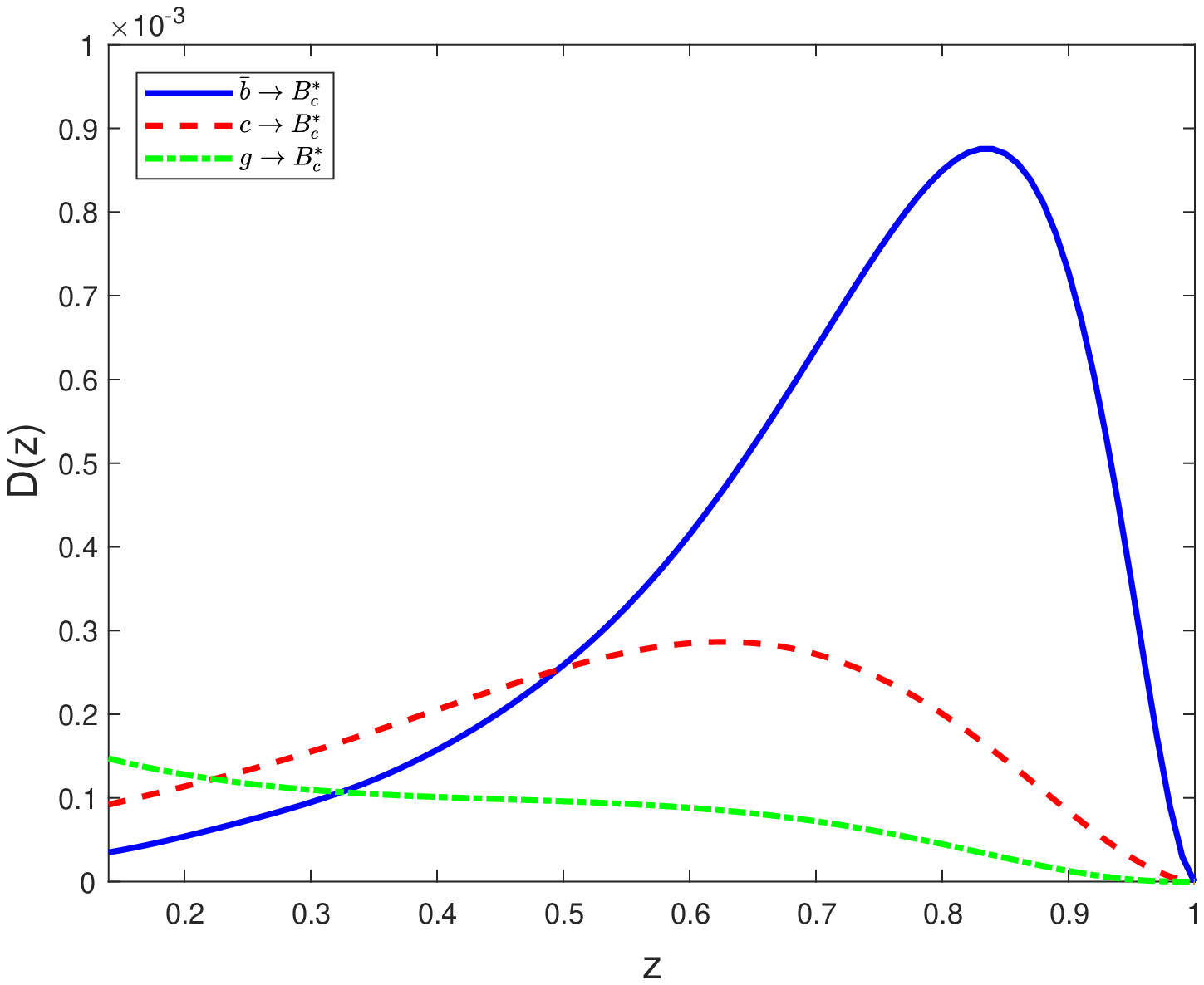}
\caption{The FFs $D_{\bar{b} \to B^*_c}(z,\mu_{F})$, $D_{c \to B^*_c}(z,\mu_{F})$ and $D_{g \to B^*_c}(z,\mu_{F})$ as functions of $z$ with $\mu_{F}=2m_b+2m_c$. In order to show these results in one figure, $D_{c \to B^*_c}(z,\mu_{F})$ and $D_{g \to B^*_c}(z,\mu_{F})$ are artificially multiplied by a factor of 30.} \label{dzmbc3s1}
\end{figure}

Although for solving the DGLAP equations, the QCD NLO FFs at comparatively low energy scale of $\bar{b}$ and $c$ quarks to the mesons $B_c$ or $B_c^*$ provide main parts of the necessary ``initial FFs", due to the mixing of the gluon’s and flavor-singlet quarks’ FFs, the FF of a gluon at the low energy scale also is a necessary part of the initial condition, so we need to calculate out the FF of a gluon at the comparatively low energy scale $\mu_F=2m_b+2m_c$, where is the threshold of the $B_c$ or $B^*_c$ production by a gluon. Now the ``initial FFs" of b̄ and c quarks as well as a gluon to the meson $B_c$ or $B^*_c$ all at the low energy scale $\mu_{F0}=2m_b+2m_c$, which as ``initial condition" are needed for solving the DGLAP equations, are shown in Figs.\ref{dzmbc1s0} and \ref{dzmbc3s1}, where the FFs of $\bar{b}$ and $c$ quarks to the meson $B_c$ or $B^*_c$ at
this energy scale are obtained by solving the DGLAP equations from $\mu_R=\mu_{F0}=m_b+2m_c$(for $\bar{b} \to B_c(B_c^*)$)
or $\mu_R=\mu_{F0}=2m_b+m_c$(for $c \to B_c(B_c^*)$). In order to
show the FF curves in one figure, in Figs.\ref{dzmbc1s0} and \ref{dzmbc3s1}, the gluon and $c$-quark FFs are artificially multiplied by a factor of 30. From Figs.\ref{dzmbc1s0} and \ref{dzmbc3s1}, one can see that the FFs for $g\to B_c(B^*_c)$
and $c \to B_c(B_c^*)$ are about 2 orders of magnitude smaller
than the FF for $\bar{b} \to B_c(B_c^*)$.

\begin{figure}[htbp]
\includegraphics[width=0.5\textwidth]{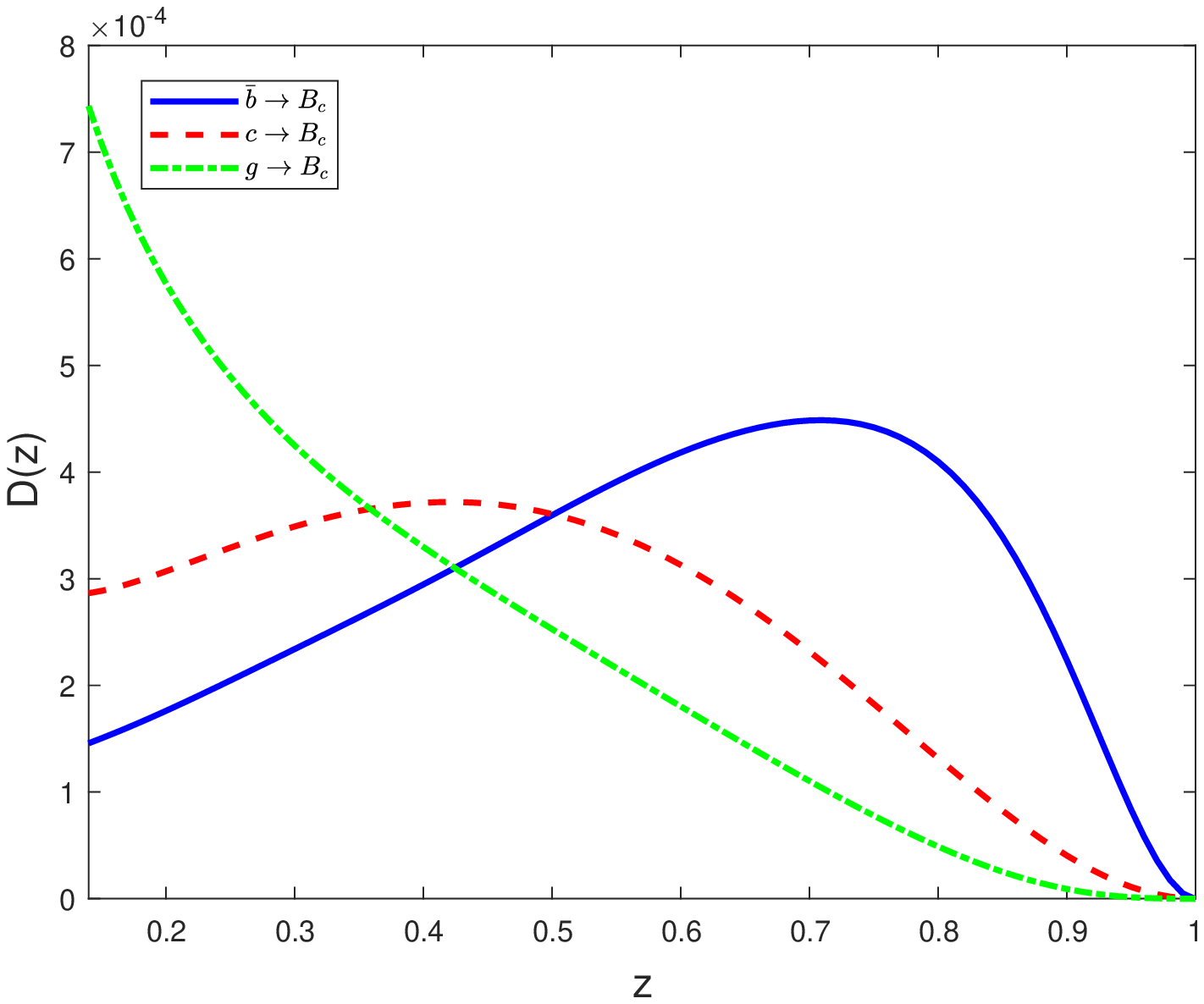}
\caption{The FFs $D_{\bar{b} \to B_c}(z,\mu_{F})$, $D_{c \to B_c}(z,\mu_{F})$ and $D_{g \to B_c}(z,\mu_{F})$ as functions of $z$ with $\mu_{F}=m_{_Z}$. In order to show these results in one figure, $D_{c \to B_c}(z,\mu_{F})$ and $D_{g \to B_c}(z,\mu_{F})$ are artificially multiplied by a factor of 30.} \label{dzmz1s0}
\end{figure}

\begin{figure}[htbp]
\includegraphics[width=0.5\textwidth]{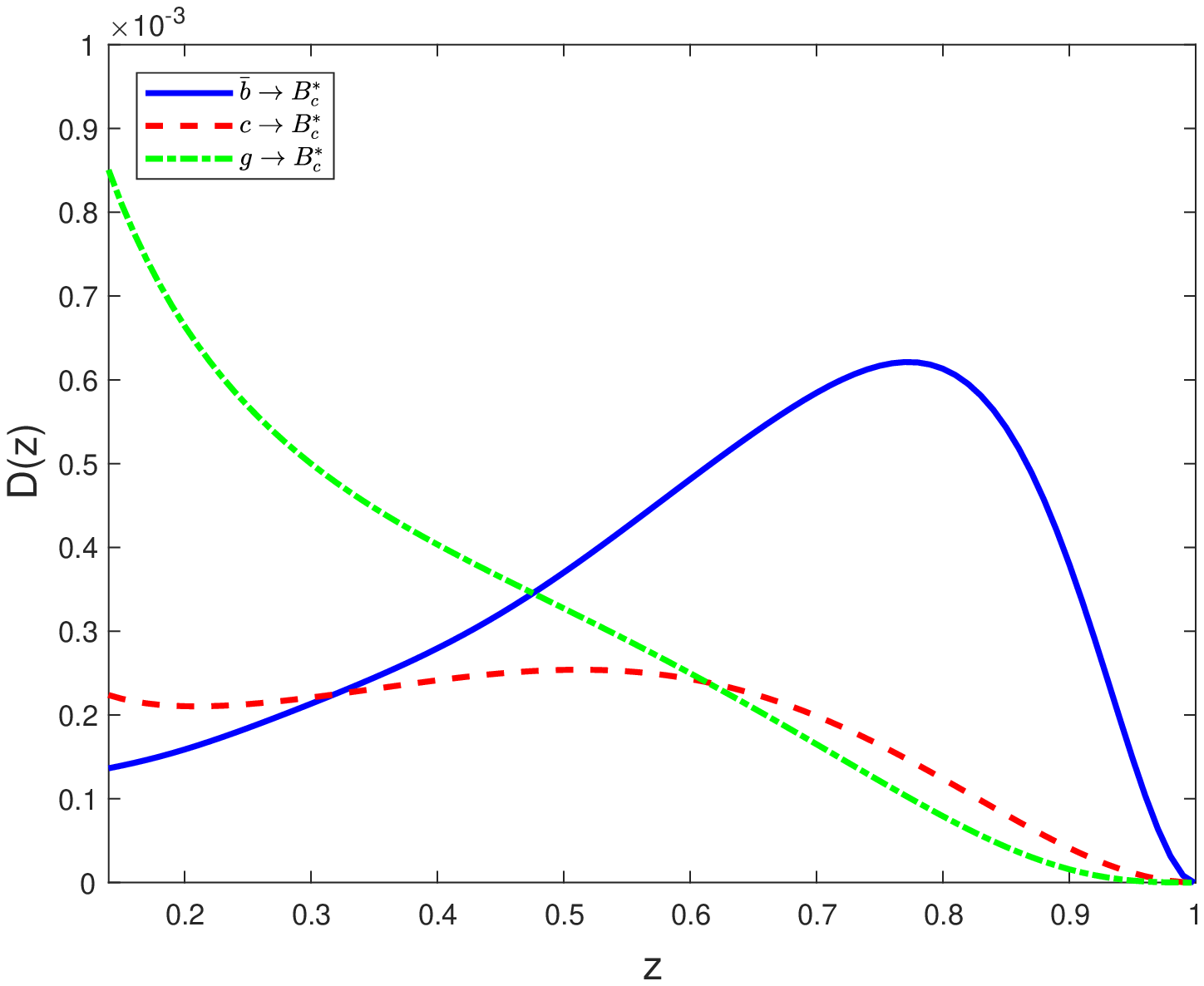}
\caption{The FFs $D_{\bar{b} \to B^*_c}(z,\mu_{F})$, $D_{c \to B^*_c}(z,\mu_{F})$ and $D_{g \to B^*_c}(z,\mu_{F})$ as functions of $z$ with $\mu_{F}=m_{_Z}$. In order to show these results in one figure, $D_{c \to B^*_c}(z,\mu_{F})$ and $D_{g \to B^*_c}(z,\mu_{F})$ are  artificially multiplied by a factor of 30.} \label{dzmz3s1}
\end{figure}

For the following application in the next section, with the
initial FFs at $\mu_F=2m_b+2m_c$ which are obtained by means of this work on the QCD NLO FFs, we calculate out the FFs at the energy scale $\mu_F=m_{_Z}$ by solving DGLAP evolution equations Eq.(\ref{eq.dglap}); and the results are shown in Figs.\ref{dzmz1s0} and \ref{dzmz3s1}. For the same reason as in Figs.\ref{dzmz1s0} and \ref{dzmz3s1}, we artificially multiply the gluon and $c$ quark FFs by a factor of 30. From Figs.\ref{dzmz1s0} and \ref{dzmz3s1}, one can see that the
FFs are changed due to the evolution. The average values of $z$ for the $\bar{b}$-quark and $c$-quark fragmentation are shifted to smaller values. For the $\bar{b}$-quark fragmentation,
\begin{eqnarray}
\langle z \rangle(B_c,\mu_F=m_{_Z})=0.58,\nonumber \\
\langle z \rangle(B^*_c,\mu_F=m_{_Z})=0.62.
\end{eqnarray}
For the $c$-quark fragmentation,
\begin{eqnarray}
\langle z \rangle(B_c,\mu_F=m_{_Z})=0.46,\nonumber \\
\langle z \rangle(B^*_c,\mu_F=m_{_Z})=0.49.
\end{eqnarray}

The fragmentation probabilities for the gluon fragmentation are increased compared to the gluon fragmentation at $\mu_F=2m_b+2m_c$. However, the fragmentation probabilities for $g\to B_c(B^*_c)$ are small compared to the fragmentation probabilities for $\bar{b} \to B_c(B^*_c)$.

\section{Application to $B_c(B_c^*)$ production at a Z factory}
\label{sec4}

The production of $B_c$ and $B_c^*$ mesons at a Z factory is the simplest case where only the fragmentation from $\bar{b}$ and $c$ quarks should be considered, and the fragmentation from light quarks and gluon (being high-order processes) can be ignored. Moreover, this production is a typical process for doubly heavy flavored hadron production at a Z factory, which can be a good reference for  doubly heavy hadron production at a Z factory. Thus, to try to have a higher accuracy for the fragmentation approach in computing the production of the $B_c$ and $B_c^*$ at a Z factory, we would like to apply the FFs, which are accurate up to the QCD NLO at a low factorization energy scale $\mu_{F0}$ and evolved with the DGLAP equations to the proper and higher energy scale
(here it is $\mu_F=m_{_Z}$ ), to computing the production of the $B_c$ and $B_c^*$ at a Z factory, and to compare the results with those obtained by the approaches of complete LO and NLO QCD.

With the pQCD factorization (\ref{eqbcfact}), the differential cross sections of $B_c $ and $B_c^*$ production at a Z factory can be calculated straightforwardly. The expressions for the coefficient functions $d\hat{\sigma}_{e^+e^-\to i+X}/dy$ ($i=\bar{b},c$) in the limit $m_Q \to 0$ can be found in Refs.\cite{coefun1,coefun2}.


For the numerical calculations, the additional and relevant input parameters are taken as follows:
\begin{eqnarray}
 \alpha=1/128 \,,\; {\sin}^2\theta_{_W}=0.231\,,\; \Gamma_{_Z}=2.4952~{\rm GeV}\,,
\end{eqnarray}
where $\alpha=\alpha(m_{_Z})$ is the electromagnetic coupling constant renormalized at $m_{_Z}$.

\begin{figure}[htbp]
\includegraphics[width=0.5\textwidth]{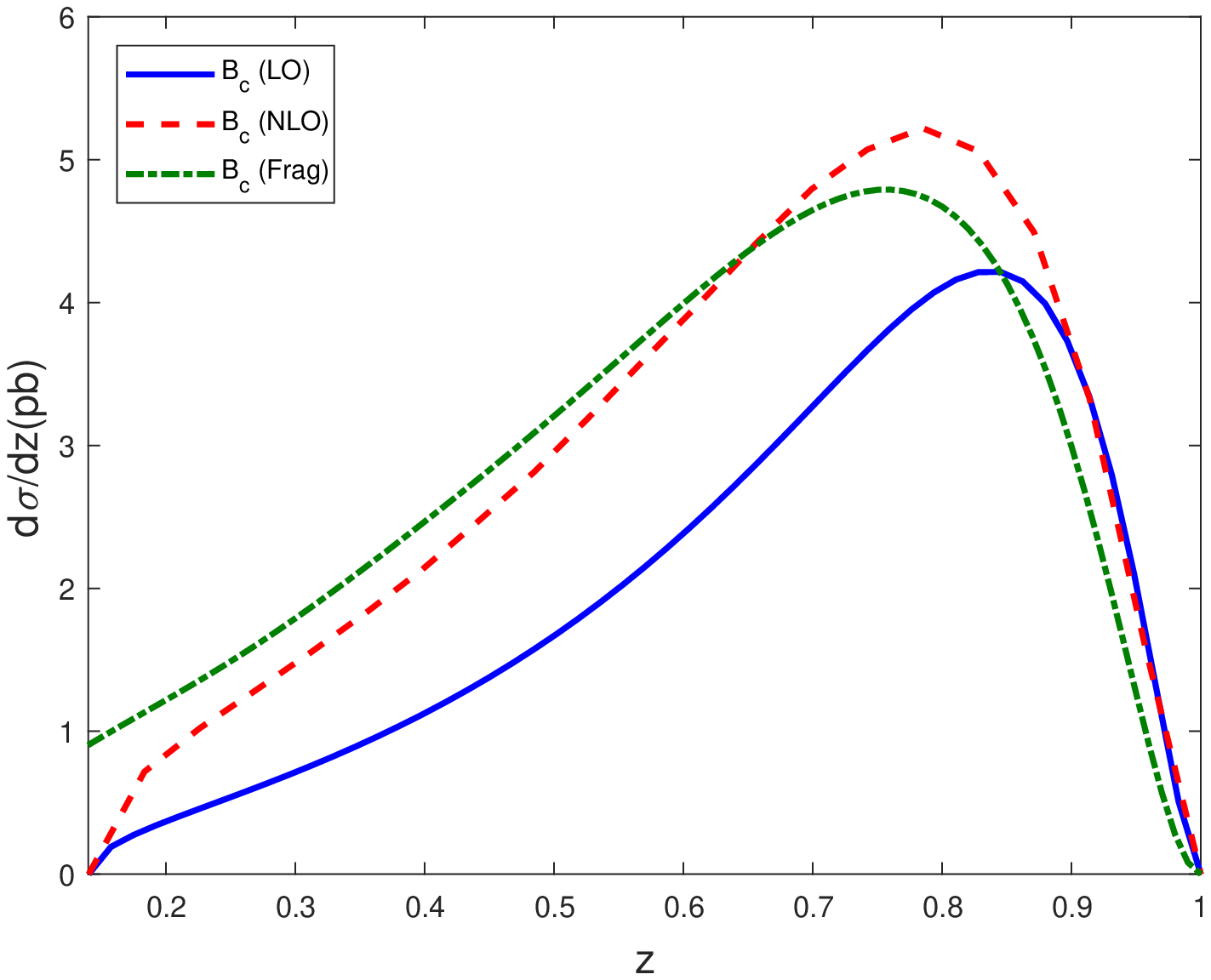}
\caption{The differential cross section $d\sigma/dz$ for the production of the $B_c$ meson as a function of $z$.} \label{sigz1}
\end{figure}

\begin{figure}[htbp]
\includegraphics[width=0.5\textwidth]{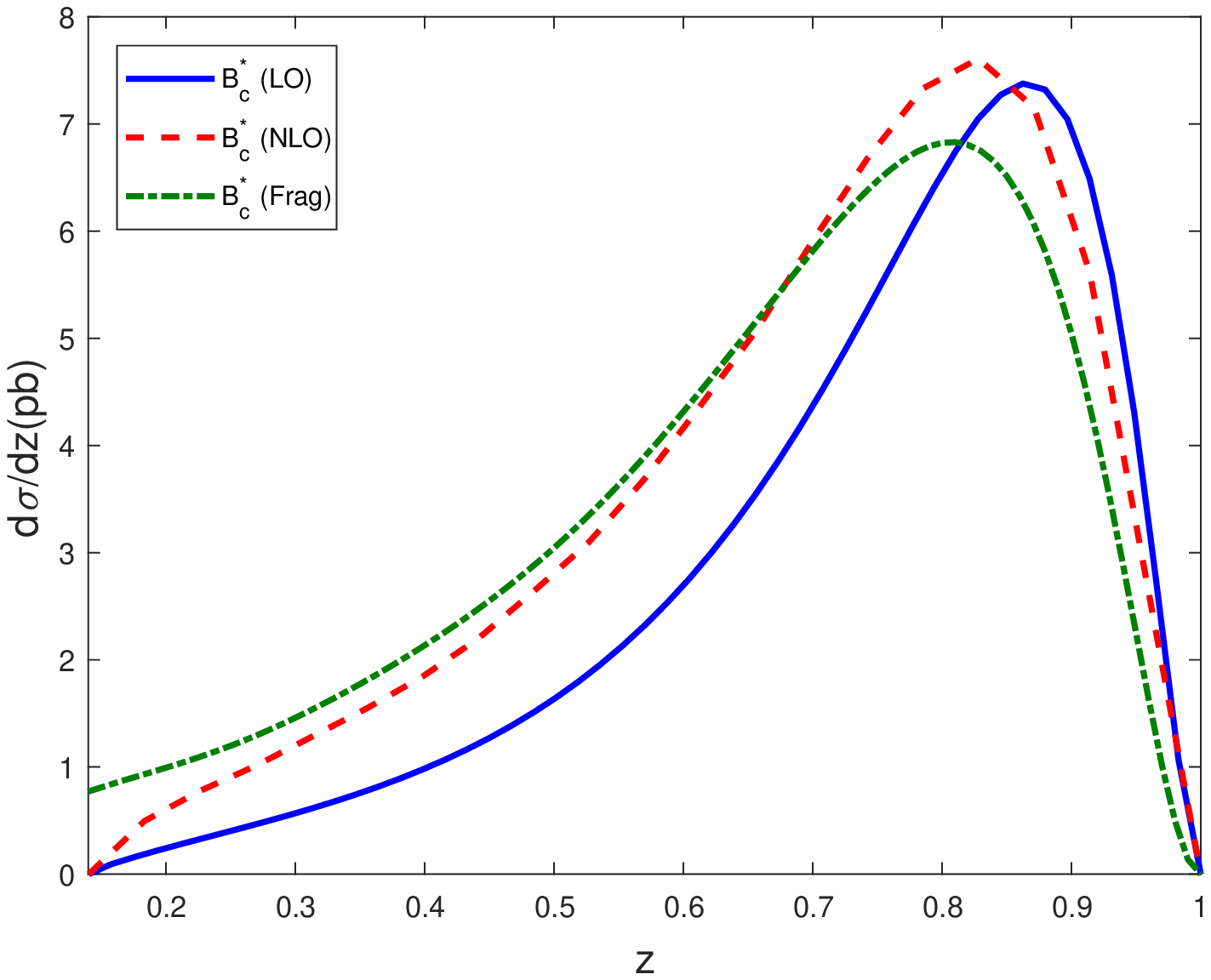}
\caption{The differential cross section $d\sigma/dz$ for the production of the $B^*_c$ meson as a function of $z$.} \label{sigz3}
\end{figure}

The differential cross sections $d\sigma/dz$ for the production of $B_c$ and $B^*_c$ mesons at the Z pole are presented in Figs.\ref{sigz1} and \ref{sigz3}, where the contributions from $\gamma$ exchange and $\gamma-Z$ interference are neglected due to the fact that they are much smaller in comparison to the contributions from Z exchange at the Z pole\cite{bcnlo}. In Figs.\ref{sigz1} and \ref{sigz3}, ``LO" and ``NLO" denote the results of the complete LO and NLO calculations, respectively, ``Frag" denotes the results of the fragmentation approach and the leading logarithms (LLs) being resummed through DGLAP evolution equations. For the results of the complete LO and NLO approaches, the renormalization is set at $\mu_R=m_b+2m_c$\footnote{In order to maintain compatibility with the results of the fragmentation calculations, for the complete NLO calculations we adopt the renormalization scale $\mu_R=m_b+2m_c$, although the complete NLO results in our previous paper\cite{bcnlo} are those with the renormalization at $\mu_R=2m_b$.}.

It is interesting to compare the total cross sections for the production of $B_c$ and $B_c^*$ mesons obtained using the fragmentation approach with those from the complete LO and NLO calculations. The obtained total cross sections are presented in Table \ref{tbccs}. We believe that the fragmentation approach provides better results for the values of the total cross sections of $B_c$ and $B_c^*$ production at the Z pole.

\begin{table}[h]
\begin{tabular}{c c c c c}
\hline\hline
~~{\rm States} ~~ &  ~~LO ~~ &  ~NLO ~& ~ Frag~  \\
\hline
$B_c$ &  1.76 & 2.53 &   2.51  \\
$B_c^*$ & 2.46 & 3.07 &  2.98  \\
\hline\hline
\end{tabular}
\caption{The total cross sections (in ${\rm pb}$) for the production of $B_c$ and $B_c^*$ mesons at the Z pole. Here, ``LO" and ``NLO" denote the results from the complete LO and NLO calculations, while ``Frag" denotes the results from the fragmentation approach.}
\label{tbccs}
\end{table}

\section{Discussions and Conclusion}
\label{conclusion}

In this paper, by means of the general operator definition of the FFs, we have derived the FFs for a $\bar{b}$ or $c$ quark fragmenting to $B_c$ and $B^*_c$ mesons in LO and NLO QCD, and the numerical results with reasonable input parameters are presented in figures.

In the derivation of the NLO ``real corrections" to the FFs, the difficulty in extracting the singularities is overcome by the fact that certain proper subtraction terms are constructed, which contain the exact same singularities ($1/\epsilon$) as those in the real corrections under dimensional regularization, but they can be computed almost analytically [see Eqs.(\ref{dsub}), (\ref{dsub1}), and (\ref{dsub2})]. Then, with the constructed auxiliary terms for subtractions, the singular and finite contributions from the real corrections can be computed separately and the finite contributions can be calculated numerically. Note that here the integrations of the subtraction terms over the phase space are carried out under suitable parametrizations, which are very similar to those introduced in Ref.\cite{braaten} , and the expressions for the subtraction terms and the phase-space parametrizations may be useful in calculating the real QCD NLO corrections for other FFs.

It is known that the choices for the factorization scale $\mu_F$ and the renormalization scale $\mu_R$ are very important in QCD calculations. For NLO corrections of FFs, one may set them equal to each other or different from each other according to convenience. As a typical case, here we set the ``(initial) factorization scale" to $\mu_{F0}=m_b+2m_c$ for the QCD NLO ``initial FFs", and the results show that the NLO corrections are significant with two possible choices of the renormalization scale. Moreover, for an important application specifically discussed in this paper, to gain a higher accuracy FFs with the
factorization $\mu_{F}=m_{_Z}$ were used. We obtained FFs by solving the DGLAP evolution equation, starting with the ``initial QCD NLO FFs" at a low energy scale $\mu_F=2m_b+2m_c$. Since the solution of the DGLAP evolution equation shows certain shifts of the average value of the energy fraction $z$ in a small region, we hope that future experiments can test this effect(s).

Finally, the production of $B_c$ and $B_c^*$ mesons at a Z factory is the simplest case where only the fragmentation from a $\bar{b}$ or $c$ quark should be considered, and this production is a typical process for doubly heavy hadron production at a Z factory, which can be used as a reference to estimate doubly heavy hadron production at a Z factory. Thus we applied the FFs at the energy scale $\mu_{F}=m_{_Z}$, which were obtained by evolving the QCD NLO ones at a low energy scale $\mu_{F0}$ and shown in Figs.\ref{dzmz1s0} and \ref{dzmz3s1}, to computing the production of $B_c$ and $B_c^*$ mesons at a Z factory, and we suspect that the results presented in Figs.\ref{sigz1} are \ref{sigz3} are comparatively accurate. For comparison, the results from the fully complete LO and NLO calculations were also presented in these figures.

In summary, we derived the QCD NLO FFs of $\bar{b}$ and $c$ quarks to $B_c$ and $B_c^*$ mesons, and the physical picture for the production of $B_c$ and $B_c^*$ mesons to QCD leading logarithm (LL) order at a Z factory was described as follows: the $\bar{b}$ and $c$ quarks are produced at high energy ($\sqrt{s}=m_{_Z}$), then the produced $\bar{b}$ and $c$ quarks are evolved to the lower invariant mass (${\cal O}(m_Q)$) by emitting real and virtual collinear gluons and quarks (that are summed by the LO DGLP equations), at last they fragment into the meson $B_c$ or $B_c^*$, that is described by the QCD NLO FFs. Therefore one may reasonably understand why the physics picture summarized here has more solid QCD foundation and works better in estimating the $B_c$ and $B_c^*$ production at a Z factory.

\vspace{4mm}

\noindent {\bf\Large Acknowledgments:} This work was supported in part by the National Natural Science Foundation of China (NSFC) under Grants No. 11745006, No. 11535002, No. 11675239, No. 11821505, No. 11275036, No. 11625520, No. 11705045, and No. 11847222. It was also supported in part by the Key Research Program of Frontier Sciences, CAS, Grant No. QYZDY-SSW-SYS006.

\appendix

\section{The expressions for the coefficients $a_i$ in the LO squared amplitudes}
\label{Ap.coeai}

For the production of the $c\bar{b}[^1S_0^{[1]}]$ state, the expressions for the coefficients $a_i$ in Eq.(\ref{aborn}) are
\begin{eqnarray}
a_2=&&z^2(1-z)[-(d~r_b+2r_c-4)z+d-2]^2,\nonumber \\
a_3=&&-4z^2(1-r_b~z)\lbrace (d~r_b+2r_c-3)r_b z^2\nonumber \\
&&-[2d~r_b^2+4(2+r_b)r_c-5]z-2d~r_c\nonumber \\
&&+d+2(r_c-r_b)\rbrace, \nonumber \\
a_4=&&-16r_b r_c z^2(1-r_b~z)^2.
\end{eqnarray}
For the production of the $c\bar{b}[^3S_1^{[1]}]$ state,
\begin{eqnarray}
a_2=&&z^2(1-z)\{(1-r_b~z)^2d^3-(1-r_b~z)[(9r_c-13)z+9]d^2\nonumber \\
&&+4 [z (7 r_c ((r_c - 3) z + 2) + 3 (5 z - 7)) + 7]d \nonumber \\
&&- 4 z [7 r_c ((r_c - 4) z + 2) + 22 z - 28] - 28\},\nonumber \\
a_3=&&-4z^2  (1-r_b z) \{z [-2 (d-5) (d-2) r_c^2\nonumber \\
&&+4 ((d-8) d+13) r_c+(15-2 d) d]\nonumber \\
&&-r_b z^2 [d (-d ~r_b-7 r_c+8)+14 r_c-15]\nonumber \\
&&- d(d-7) (2 r_c-1)-20 r_c-29 z+14\}, \nonumber \\
a_4=&&-16  r_b r_c (d-1) z^2 (1-r_b z)^2.
\end{eqnarray}

\section{Phase space for the real corrections}
\label{Ap.psr}

The differential phase space for the real corrections to the FFs is
\begin{eqnarray}
d\phi_{\rm real}=&&\frac{ d^{d-1}\textbf{p}_2}{(2\pi)^{d-1}2p_2^0}\frac{ d^{d-1}\textbf{p}_3}{(2\pi)^{d-1}2p_3^0} \nonumber \\
&&\times 2\pi \delta\left(K\cdot n-(p_1+p_2+p_3)\cdot n\right).\label{eqb1}
\end{eqnarray}
The different parameterizations are required in order to extract the poles in $\epsilon$ in the real corrections. We adopt similar parametrizations as those used in Ref.\cite{braaten}. In Ref.\cite{braaten}, the authors derived the phase space for two massless partons in final state. In our case, there is one massive parton and one massless parton in the final state, so we derive the formulas for this case.

The differential phase space for a single parton with momentum $p$ and mass $m$ can be expressed as
\begin{eqnarray}
\frac{d^{d-1}\textbf{p}}{(2\pi)^{d-1}2p^0}=\frac{\vert \textbf{p}\vert^{2-2\epsilon}\vert{\rm sin} \theta \vert^{1-2\epsilon}}{2p^0(2\pi)^{3-2\epsilon}}  d\vert \textbf{p}\vert d\theta d\Omega_{\perp},\label{eqb2}
\end{eqnarray}
where $\theta$ denotes the polar angle and $d\Omega_{\perp}$ denotes the differential transverse solid angle. The total transverse solid angle $\Omega_{\perp}=\int d\Omega_{\perp}= 2\pi^{1-\epsilon}/\Gamma(1-\epsilon)$.

It is useful to introduce a light-like momentum $k$, and define the variable
\begin{eqnarray}
\lambda=2 k\cdot p/k\cdot n.\label{eqb3}
\end{eqnarray}
Then, the differential phase space for a single parton can be expressed as
\begin{eqnarray}
\frac{d^{d-1}\textbf{p}}{(2\pi)^{d-1}2p^0}=\frac{(\lambda p\cdot n-m^2)^{-\epsilon}}{4(2\pi)^{3-2\epsilon}}  d\lambda \, d (p\cdot n) \,d\Omega_{\perp}.\label{eqb4}
\end{eqnarray}
Here, $d\Omega_{\perp}$ is Lorentz invariant due to the fact that the differential phase space, $\lambda$ and $p\cdot n$ are Lorentz invariant. This expression can be easily derived from Eq.(\ref{eqb2}) in a Lorentz frame where the spatial parts of the light-like vectors $n$ and $k$ are back to back. The differential phase space for a massless parton can be easily obtained from Eq.(\ref{eqb4}) by setting $m=0$.

We can apply the parametrization Eq.(\ref{eqb4}) to the differential phase spaces for $p_2$ and $p_3$ in Eq.(\ref{eqb1}). Two light-like vectors $k_2$ and $k_3$ corresponding to the parametrizations of $p_2$ and $p_3$ are introduced. The integral over $p_2 \cdot n$ can be carried out through the $\delta$ function, and the integral over $\Omega_{2\perp}$ is trivial. Then we obtain the expression
\begin{eqnarray}
d\phi_{\rm real}=&&\frac{2^{-2\epsilon}[(1-z)K\cdot n]^{1-2\epsilon}}{(4\pi)^{4-3\epsilon}\Gamma(1-\epsilon)}[u(1-u)]^{-\epsilon} \lambda_2^{-\epsilon}\lambda_3^{-\epsilon}\nonumber \\
&&\left[1-\frac{m_c^2}{\lambda_2(1-u)(1-z)K\cdot n}\right]^{-\epsilon}du\, d\lambda_2 \,d\lambda_3 \, d\Omega_{3\perp}, \nonumber \\ \label{eqb5}
\end{eqnarray}
where
\begin{eqnarray}
\lambda_2=2 k_2\cdot p_2/k_2\cdot n,~~~~\lambda_3=2 k_3\cdot p_3/k_3\cdot n.\label{eqb6}
\end{eqnarray}
We have converted the integral variable $p_3 \cdot n$ to $u$ by using the definition of $u$ in Eq.(\ref{defvar}). If we set $m_c=0$ in Eq.(\ref{eqb5}), we obtain an expression that is the same as Eq.(A.6) in Ref.\cite{braaten}.

We need to choose proper light-like vectors $k_2$ and $k_3$ in order to extract the poles in $\epsilon$. For the subtraction terms that contain $s$, we choose
\begin{eqnarray}
k_2^{\mu}&=&p_1^{\mu}-\frac{M^2}{2p_1\cdot n}n^{\mu},  \nonumber \\
k_3^{\mu}&=&(p_1+p_2)^{\mu}-\frac{s_1}{2(p_1+p_2)\cdot n}n^{\mu},
\label{eqb7}
\end{eqnarray}
then
\begin{eqnarray}
\lambda_2=\frac{1}{zK\cdot n}\left(s_1-m_c^2-\frac{1-u+uz}{z}M^2\right),\label{eqb8}
\end{eqnarray}
and
\begin{eqnarray}
\lambda_3=\frac{1}{(1-u+uz)K\cdot n}\left(s-\frac{s_1}{1-u+uz}\right),\label{eqb9}
\end{eqnarray}
Changing variables in Eq.(\ref{eqb5}) from $u$, $\lambda_2$ and $\lambda_3$ to $y$, $s_1$ and $s$, we obtain
\begin{eqnarray}
d\phi_{\rm real}=&&\frac{2^{-2\epsilon}z^{-1+\epsilon}}{(4\pi)^{4-3\epsilon}\Gamma(1-\epsilon)K\cdot n}y^{-1+\epsilon}(1-y)^{-\epsilon}(y-z)^{-\epsilon} \nonumber \\
&&\times (s-s_1/y)^{-\epsilon}[s_1-m_c^2-M^2/(z/y)]^{-\epsilon} \nonumber \\
&&\times \left[1-\frac{zm_c^2}{(y-z)(s_1-m_c^2-y M^2/z)}\right]^{-\epsilon}dy\, ds \,ds_1 \, d\Omega_{3\perp}. \nonumber \\ \label{eqb10}
\end{eqnarray}

For the subtraction terms that contain $s_2$, we choose
\begin{eqnarray}
k_2^{\mu}=p_1^{\mu}-\frac{M^2}{2p_1\cdot n}n^{\mu},   k_3^{\mu}=p_{12}^{\mu}-\frac{m_b^2}{2p_{12}\cdot n}n^{\mu}.\label{eqb11}
\end{eqnarray}
Then
\begin{eqnarray}
&&\lambda_2=\frac{1}{zK\cdot n}\left(s_1-m_c^2-\frac{1-u+uz}{z}M^2\right),\nonumber \\
&&\lambda_3=\frac{1}{r_b z K\cdot n}\left(s_2-\frac{r_b z+u(1-z)}{r_bz}m_b^2\right),\label{eqb12}
\end{eqnarray}
After changing variables in Eq.(\ref{eqb5}) from $u$, $\lambda_2$ and $\lambda_3$ to $y$, $s_1$ and $s_2$, we obtain
\begin{eqnarray}
d\phi_{\rm real}=&&\frac{2^{-2\epsilon}(r_b z^2)^{-1+\epsilon}}{(4\pi)^{4-3\epsilon}\Gamma(1-\epsilon)K\cdot n}(1-y)^{-\epsilon}(y-z)^{-\epsilon} \nonumber \\
&&\times \left(s_2-\frac{1-y+r_bz}{r_bz}m_b^2\right)^{-\epsilon}[s_1-m_c^2-M^2/(z/y)]^{-\epsilon} \nonumber \\
&&\times \left[1-\frac{zm_c^2}{(y-z)(s_1-m_c^2-y M^2/z)}\right]^{-\epsilon}dy\, ds_1 \,ds_2 \, d\Omega_{3\perp}. \nonumber \\ \label{eqb13}
\end{eqnarray}

For the subtraction terms that contain $s_3$, we choose
\begin{eqnarray}
&&k_2^{\mu}=p_1^{\mu}-\frac{M^2}{2p_1\cdot n}n^{\mu},  \nonumber \\ &&k_3^{\mu}=(p_{11}+p_2)^{\mu}-\frac{(p_{11}+p_2)^2}{2(p_{11}+p_2)\cdot n}n^{\mu}.\label{eqb14}
\end{eqnarray}
Then,
\begin{eqnarray}
&&\lambda_2=\frac{1}{zK\cdot n}\left(s_1-m_c^2-\frac{1-u+uz}{z}M^2\right),\nonumber \\
&&\lambda_3=\frac{1}{(y-r_bz) K\cdot n}\left[s_3-\frac{r_c(1-r_b z)(s_1-m_b^2)}{y-r_bz}\right],\label{eqb15}
\end{eqnarray}
After changing variables in Eq.(\ref{eqb5}) from $u$, $\lambda_2$ and $\lambda_3$ to $y$, $s_1$ and $s_3$, we obtain
\begin{eqnarray}
d\phi_{\rm real}=&&\frac{2^{-2\epsilon}z^{-1+\epsilon}}{(4\pi)^{4-3\epsilon}\Gamma(1-\epsilon)K\cdot n}(1-y)^{-\epsilon}(y-z)^{-\epsilon} \nonumber \\
&&\times (y-r_bz)^{-1+\epsilon}\left[s_3-\frac{r_c(1-r_bz)(s_1-m_b^2)}{y-r_bz}\right]^{-\epsilon}\nonumber \\
&&\times[s_1-m_c^2-M^2/(z/y)]^{-\epsilon} \nonumber \\
&&\times \left[1-\frac{zm_c^2}{(y-z)(s_1-m_c^2-y M^2/z)}\right]^{-\epsilon}dy\, ds_1 \,ds_3 \, d\Omega_{3\perp}. \nonumber \\ \label{eqb16}
\end{eqnarray}

To derive the differential phase space for the subtraction terms that contain $t_1$ or $t_2$, we multiply Eq.(\ref{eqb1}) by
\begin{eqnarray}
 \int_0^{\infty}dt_2 \int d^d\tilde{p}~ && \delta^d\left(\tilde{p}-p_2-p_3+\frac{t_2}{2(p_2+p_3)\cdot n}\right)\nonumber \\
&&\times \delta(t_2-2p_2\cdot p_3),\label{eqb17}
\end{eqnarray}
which is equal to 1 and does not change the phase space. After integrating over $p_2$, the differential phase space can be expressed as
\begin{eqnarray}
d\phi_{\rm real}=&&\frac{ d^{d-1}\tilde{\textbf{p}}}{(2\pi)^{d-1}2\tilde{p}^0}\frac{ d^{d-1}\textbf{p}_3}{(2\pi)^{d-1}2p_3^0} \frac{\tilde{p}\cdot n}{(\tilde{p}-p_3)\cdot n}\nonumber \\
&&\times 2\pi \delta\left(K\cdot n-(p_1+\tilde{p})\cdot n\right).\label{eqb18}
\end{eqnarray}
Using the parametrization Eq.(\ref{eqb4}) on the differential phase spaces for $\tilde{p}$ and $p_3$ in Eq.(\ref{eqb18}), we obtain the expression
\begin{eqnarray}
d\phi_{\rm real}=&&\frac{2^{-2\epsilon}[(1-z)K\cdot n]^{1-2\epsilon}}{(4\pi)^{4-3\epsilon}\Gamma(1-\epsilon)}\frac{u^{-\epsilon}}{1-u} \tilde{\lambda}^{-\epsilon}\lambda_3^{-\epsilon}\nonumber \\
&&\left[1-\frac{m_c^2}{\tilde{\lambda}(1-z)K\cdot n}\right]^{-\epsilon}du\, d\tilde{\lambda} \,d\lambda_3 \, d\Omega_{3\perp}, \nonumber \\ \label{eqb19}
\end{eqnarray}
where
\begin{eqnarray}
\tilde{\lambda}=2 \tilde{k}\cdot \tilde{p}/\tilde{k}\cdot n,~~~~\lambda_3=2 k_3\cdot p_3/k_3\cdot n.\label{eqb20}
\end{eqnarray}

For the subtraction terms that contain $t_1$, we choose the light-like vectors $\tilde{k}$ and $k_3$ as follows
\begin{eqnarray}
\tilde{k}^{\mu}=p_1^{\mu}-\frac{M^2}{2p_1\cdot n}n^{\mu},\,k_3^{\mu}=\tilde{k}^{\mu},\label{eqb21}
\end{eqnarray}
then
\begin{eqnarray}
&&\tilde{\lambda}=\frac{1}{zK\cdot n}\left(\tilde{s}-m_c^2-M^2/z\right),\nonumber \\
&&\lambda_3=\frac{1}{z K\cdot n}\left[t_1-(1/z-1)M^2 u\right],\label{eqb22}
\end{eqnarray}
After changing variables in Eq.(\ref{eqb19}) from $\tilde{\lambda}$ and $\lambda_3$ to  $\tilde{s}$ and $t_1$, we obtain
\begin{eqnarray}
d\phi_{\rm real}=&&\frac{2^{-2\epsilon}z^{-2+2\epsilon}(1-z)^{1-2\epsilon}}{(4\pi)^{4-3\epsilon}\Gamma(1-\epsilon)K\cdot n}\frac{u^{-\epsilon}}{1-u} \nonumber \\
&&\times[t_1-(1/z-1)M^2 u]^{-\epsilon}[\tilde{s}-m_c^2-M^2/z]^{-\epsilon}  \nonumber \\
&&\times \left[1-\frac{zm_c^2}{(1-z)(\tilde{s}-m_c^2-M^2/z)}\right]^{-\epsilon} du\, d\tilde{s} \,dt_1 d\Omega_{3\perp}. \nonumber \\ \label{eqb23}
\end{eqnarray}

For the subtraction terms that contain $t_2$, we choose the light-like vectors $\tilde{k}$ and $k_3$ as follows:
$$\tilde{k}^{\mu}=p_1^{\mu}-\frac{M^2}{2p_1\cdot n}n^{\mu},  \;\;\;\; k_3^{\mu}=\tilde{p}^{\mu}-\frac{m_c^2}{2\tilde{p}\cdot n}n^{\mu},$$
then
$$\tilde{\lambda}=\frac{1}{zK\cdot n}\left(\tilde{s}-m_c^2-M^2/z\right),$$
$$ \lambda_3=\frac{1}{(1-z) K\cdot n}\left[(1-u)t_2-m_c^2 u\right].$$
After changing variables from $\tilde{\lambda}$ and $\lambda_3$ to $\tilde{s}$ and $t$, we obtain
\begin{eqnarray}
d\phi_{\rm real}=&&\frac{2^{-2\epsilon}z^{-1+\epsilon}(1-z)^{-\epsilon}}{(4\pi)^{4-3\epsilon}\Gamma(1-\epsilon)K\cdot n}u^{-\epsilon} \nonumber \\
&&\times[(1-u)t_2-m_c^2 u]^{-\epsilon}[\tilde{s}-m_c^2-M^2/z]^{-\epsilon}  \nonumber \\
&&\times \left[1-\frac{zm_c^2}{(1-z)(\tilde{s}-m_c^2-M^2/z)}\right]^{-\epsilon} du\, d\tilde{s} \,dt_2 \, d\Omega_{3\perp}. \nonumber \\ \label{eqb26}
\end{eqnarray}

\end{document}